\documentclass{article}

\usepackage{PRIMEarxiv}

\usepackage[utf8]{inputenc} % allow utf-8 input
\usepackage[T1]{fontenc}    % use 8-bit T1 fonts
\usepackage{hyperref}       % hyperlinks
\usepackage{url}            % simple URL typesetting
\usepackage{booktabs}       % professional-quality tables
\usepackage{amsfonts}       % blackboard math symbols
\usepackage{nicefrac}       % compact symbols for 1/2, etc.
\usepackage{microtype}      % microtypography
\usepackage{lipsum}
\usepackage{fancyhdr}       % header
\usepackage{graphicx}       % graphics
\graphicspath{{media/}}     % organize your images and other figures under media/ folder

\usepackage{graphicx}
\usepackage{mathtools}
\usepackage{amsfonts}
\usepackage{amsmath}
\usepackage{calrsfs}
\DeclareMathAlphabet{\mathcal}{OMS}{cmsy}{m}{n}

\usepackage{booktabs}
\usepackage{breqn}
\usepackage{xcolor}
\usepackage{hyperref}

\usepackage{float}

\usepackage{rotating}

\usepackage[utf8]{inputenc}

\usepackage[version=4]{mhchem}
\usepackage{siunitx}
\usepackage{longtable,tabularx}
\usepackage{bm}
\usepackage[inline]{enumitem}
\usepackage{array}

\newcommand{\eg}{{\em e.g.}}
\newcommand{\etal}{{\em et~al.}}
\newcommand{\ie}{{\em i.e.}}

\newcommand{\RNum}[1]{\uppercase\expandafter{\romannumeral #1\relax}}

\usepackage[linesnumbered,ruled,vlined]{algorithm2e} 

\title{GAN-DUF: Hierarchical Deep Generative Models for Design Under Free-Form Geometric Uncertainty}

\author{
  Wei (Wayne) Chen, Doksoo Lee, Oluwaseyi Balogun, Wei Chen \\
  Department of Mechanical Engineering\\
  Northwestern University\\
  Evanston, IL 60208\\
  \texttt{\{wei.wayne.chen, dslee, o-balogun, weichen\}@northwestern.edu} \\
}

\begin{document}
\maketitle

\begin{abstract}
Deep generative models have demonstrated effectiveness in learning compact and expressive design representations that significantly improve geometric design optimization. However, these models do not consider the uncertainty introduced by manufacturing or fabrication. Past work that quantifies such uncertainty often makes simplifying assumptions on geometric variations, while the ``real-world", ``free-form" uncertainty and its impact on design performance are difficult to quantify due to the high dimensionality. To address this issue, we propose a Generative Adversarial Network-based Design under Uncertainty Framework (GAN-DUF), which contains a deep generative model that simultaneously learns a compact representation of nominal (ideal) designs and the conditional distribution of fabricated designs given any nominal design. This opens up new possibilities of 1)~building a universal uncertainty quantification model compatible with both shape and topological designs, 2)~modeling free-form geometric uncertainties without the need to make any assumptions on the distribution of geometric variability, and 3)~allowing fast prediction of uncertainties for new nominal designs. We can combine the proposed deep generative model with robust design optimization or reliability-based design optimization for design under uncertainty. We demonstrated the framework on two real-world engineering design examples and showed its capability of finding the solution that possesses better performance after fabrication.
\end{abstract}

% keywords can be removed

%%%%%%%%%%%%%%%%%%%%%%%%%%%%%%%%%%%%%%%%%%%%%%%%%%%%%%%%%%%%%%%%%%%%%%
\section{INTRODUCTION}

Many engineering design problems boil down to geometric optimization. However, geometric optimization remains a grand challenge because of its extreme dimensional complexity and often hard-to-achieve performance objective. Recent work has shown that deep generative models can learn a compact\footnote{Here ``compact" means (1)~the learned design representation has a much lower dimensionality than the original representation and (2)~the learned representation contains mostly valid designs.} and expressive design representation that remarkably improves geometric design optimization performances (indicated by both the quality of optimal solutions and the computational cost)~\cite{chen2020airfoil,chen2021deep,chen2021mo}. However, past work based on deep generative models only considers the ideal scenario where manufacturing or fabrication imperfections do not occur, which is unrealistic due to the existence of uncertainties in reality, such as limited tool precision or wear. Such imperfections sometimes have a high impact on a design's performance or properties. Consequently, the originally optimal solution (obtained by only considering the ideal scenario) might not possess high performance or desired properties after fabrication.

Past work has developed task-specific robust optimization techniques to identify geometric design solutions that are insensitive to variations of load, materials, and geometry~\cite{chen2010level,chen2011new,wang2019robust}. However, due to the lack of generalizable uncertainty representation that is compatible with different geometric representations, previous work often makes simplifying assumptions on geometric variations (\eg, the distribution or the upper/lower bound of uncertain parameters), while the ``real-world", ``free-form" geometric uncertainty and its impact on design performance are difficult to quantify due to the high dimensionality\footnote{Due to the free-form nature of the geometry and its uncertainty, there is no easy way to parameterize them using lower-dimensional parameters. This is the reason for their high dimensionality.}. In this paper, we propose a \textit{Generative Adversarial Network-based Design under Uncertainty Framework (GAN-DUF)} to allow uncertainty quantification (UQ) of free-form geometric variability under real-world scenarios. The term ``free-form" refers to two aspects: 1)~the geometric variability has no shape or topological restrictions and 2)~no assumption on the form of uncertainty is needed. Therefore, this framework is generalizable to any shape or topological designs. It improves existing geometric design under uncertainty from four aspects: 1)~The generative adversarial network (GAN) uses a compact representation to reparameterize geometric designs, allowing accelerated optimization; 2)~The GAN associates real-world, free-form fabrication uncertainty with ideal designs (\ie, \textit{nominal designs}) by learning a conditional distribution of fabricated designs given any nominal design; 3)~The optimization process accounts for the distribution of geometric variability underlying any manufacturing processes, and allows UQ for robust design optimization or reliability-based design optimization; 4)~The compact representation of nominal designs allows gradient-free global optimization due to the representation's low-dimensionality.

We list the contributions of this work as follows:
\begin{enumerate}
    \item We propose a hierarchical deep generative model to simultaneously learn a compact representation of designs and quantify their real-world, free-form geometric uncertainties.
    \item We combine the proposed model with a robust design optimization framework and demonstrate its effectiveness on two realistic robust design examples.
    \item We build two benchmark datasets, containing nominal and fabricated designs, which will facilitate future studies on data-driven design under manufacturing uncertainty.
\end{enumerate}

The rest of the paper is organized as follows: Section~\ref{sec:background} introduces two core concepts used in our current work~\textemdash~the GAN and design under uncertainty, including a brief review of related past work on design under geometric uncertainty. Section~\ref{sec:methodology} describes the method in detail. Section~\ref{sec:results} demonstrates the effectiveness of the proposed method by using two case studies. Section~\ref{sec:conclusion} concludes the paper and discusses future work.

%%%%%%%%%%%%%%%%%%%%%%%%%%%%%%%%%%%%%%%%%%%%%%%%%%%%%%%%%%%%%%%%%%%%%%
\section{BACKGROUND}
\label{sec:background}

In this section, we introduce Generative Adversarial Networks and design under uncertainty.

\subsection{Generative Adversarial Networks}
%$\partial \Omega^*$
%$\partial \Omega(\pmb{\xi})$
The generative adversarial network~\cite{goodfellow2014generative} models a game between a \textit{generator} $G$ and a \textit{discriminator} $D$. The goal of $G$ is to generate samples (designs in our case) that resemble those from data; while $D$ tries to distinguish between real data and generated samples by predicting the probability of a sample being real data. Both models improve during training via the following minimax optimization:
\begin{equation}
\begin{split}
\min_G\max_D \mathbb{E}_{\mathbf{x}\sim P_{\text{data}}}[\log D(\mathbf{x})] + \mathbb{E}_{\mathbf{z}\sim P_{\mathbf{z}}}[\log(1-D(G(\mathbf{z})))],
\label{eq:gan_loss}
\end{split}
\end{equation}
where $\mathbb{E}$ denotes expectation, $P_{\text{data}}$ is the data distribution, $P_{\mathbf{z}}$ is a predefined noise distribution (\eg, a standard normal distribution), $\mathbf{z}\sim P_{\mathbf{z}}$ is the noise that serves as $G$'s input, $G(\mathbf{z})$ denotes the generated sample, and $\mathbf{x}$ denotes the sample from a training dataset. In this paper, $\mathbf{x}$ represents geometric designs (\eg, pixelated, voxelated, point cloud, point sequence, or surface mesh representation, usually with high-dimensionality). A trained generator thus can map from a predefined noise distribution to the distribution of designs. Due to the lower dimensionality of $\mathbf{z}$ compared to the original design representation $\mathbf{x}$, we can use $\mathbf{z}$ to more efficiently control the geometric variation of high-dimensional designs.

Despite the ability to generate high-dimensional data from low-dimensional noise, standard GANs do not have a way of regularizing the noise; so it usually cannot reflect an intuitive design variation, which is unfavorable in many design applications. To compensate for this weakness, the InfoGAN adds the \textit{latent codes} $\mathbf{c}$ as $G$'s another input and maximizes the mutual information between $\mathbf{c}$ and $G(\mathbf{c},\mathbf{z})$, as a way of regularizing $\mathbf{c}$ and obtaining an interpretable and disentangled latent representation~\cite{chen2016infogan}. The generated design is now expressed as $G(\mathbf{c},\mathbf{z})$. Maximizing the mutual information prevents the information loss of $\mathbf{c}$ in the generation process. However, directly maximizing the mutual information is difficult, the InfoGAN instead maximizes the mutual information lower bound $L_I$:
\begin{equation}
L_I(G,Q) = \mathbb{E}_{\mathbf{c}\sim P(\mathbf{c}),\mathbf{x}\sim G(\mathbf{c},\mathbf{z})}[\log Q(\mathbf{c}|\mathbf{x})] + H(\mathbf{c}),
\label{eq:li}
\end{equation}
where $H(\mathbf{c})$ is the entropy of the latent codes, and $Q$ is the auxiliary distribution for approximating $P(\mathbf{c}|\mathbf{x})$. The InfoGAN's training objective becomes:
\begin{equation}
\begin{split}
\min_{G,Q}\max_D \mathbb{E}_{\mathbf{x}\sim P_{\text{data}}}[\log D(\mathbf{x})] + \\ \mathbb{E}_{\mathbf{c}\sim P_{\mathbf{c}},\mathbf{z}\sim P_{\mathbf{z}}}[\log(1-D(G(\mathbf{c},\mathbf{z})))] - \lambda L_I(G,Q),
\end{split}
\label{eq:infogan}
\end{equation}
where $\lambda$ is a weight parameter. In practice, $H(\mathbf{c})$ is usually treated as a constant as $P_{\mathbf{c}}$ is fixed. Please see \cite{chen2016infogan} for more details about InfoGAN.

Past works studied how the maximization of $L_I$ in InfoGAN affects the latent representation of designs~\cite{chen2020airfoil,chen2019synthesizing}.
Specifically, Ref.~\cite{chen2020airfoil} showed that by maximizing the mutual information lower bound between the latent vector and the generated designs, the latent vector $\mathbf{c}$ encoded major geometric variation in the data, while the noise vector $\mathbf{z}$ encoded minor geometric variation. Ref.~\cite{chen2019synthesizing} showed that without maximizing $L_I$, the latent vector $\mathbf{c}$ either failed to capture major geometric variation or led to inconsistent geometric variation along latent space bases. The ability to capture major geometric variation is necessary, since the generator needs to cover the complete range of both the nominal and the fabricated designs. The property of consistent geometric variation is also essential for designers to explore the latent space in an intuitive way and disentangle the objective function of a design problem.

Since InfoGAN provides an interpretable and disentangled latent representation that is also compact and low-dimensional, searching for design solutions in this latent space is much more efficient than searching in the original high-dimensional design space~\cite{chen2019synthesizing,chen2020airfoil,chen2021deep,chen2021mo}. 
Building on the InfoGAN model, this work proposes a new deep generative model that constructs a hierarchical latent representation~\textemdash~the \textit{parent latent representation} encodes nominal design variation and the \textit{child latent representation} encodes fabricated design variation. In this way, we can simultaneously model 1)~the compact latent representation of nominal designs and 2)~the distribution of fabricated designs conditioned on any nominal design. We will elaborate on our proposed hierarchical generative model in Sec.~\ref{sec:method_gan}.

\subsection{Design under Uncertainty}

Design under uncertainty aims to identify optimal designs that are robust and/or reliable under the variations associated with various sources of uncertainties (\eg, material, geometry, and operating conditions)~\cite{du2004integrated, maute2014topology}. Two common approaches are robust design optimization (RDO)~\cite{chen1996procedure} and reliability-based design optimization (RBDO)~\cite{du2004sequential, choi2007reliability}.

We assume that the design is represented by a vector of deterministic design variables $\mathbf{b}$ and a vector of random variables $\pmb{\xi}$. The vector $\pmb{\xi}$ represents the sources of uncertainty (\eg, noise or control factor). The goal of RDO is to minimize the effects of variation without eliminating the sources of uncertainties~\cite{chen1996procedure}. Given a performance function $f(\cdot)$, RDO approaches simultaneously maximize the mean performance $\mu(f(\mathbf{b}, \pmb{\xi}))$ and minimize the variance of the cost $\sigma^2(f(\mathbf{b}, \pmb{\xi}))$ over $\pmb{\xi}$. The design goal, in general, involves the following optimization problem:%~\cite{chen2011new}:
\begin{equation}
\min_{\mathbf{b}} J_{\text{RDO}}( \mathbf{b}, \pmb{\xi})= -F (\mu(f(\mathbf{b}, \pmb{\xi})), \sigma(f(\mathbf{b}, \pmb{\xi}))),
\label{eq:RDO}
\end{equation}
where $F$ is the multi-objective cost function that is typically formulated as
\begin{equation}
    F(\cdot)=\mu(\cdot)-k\sigma(\cdot),
    \label{eq:F}
\end{equation}
where $k>0$ is the tuning parameter.
$\mu(f(\cdot))$ and $ \sigma^2(f(\cdot))$ are the statistical moments of $f(\cdot)$ w.r.t. the associated uncertainty $\pmb{\xi}$ and can be expressed as:
\begin{equation}
   \begin{split}
    \mu( f( \mathbf{b}, \pmb{\xi}))
    &=\mathbb{E}_{\pmb{\xi}} [ f( \mathbf{b}, \pmb{\xi})]
    =\int_{\pmb{\xi}} p(\pmb{\xi}) f( \mathbf{b}, \pmb{\xi}) \,
    d{\pmb{\xi}}, \ 
    \\
    \sigma^2( f( \mathbf{b}, \pmb{\xi}) )
    &=\mathbb{E}_{\pmb{\xi}} \left[ (f(\mathbf{b}, \pmb{\xi}) - \mu(f(\mathbf{b}, \pmb{\xi})))^2 \right]\\
    &=\int_{\pmb{\xi}} p(\pmb{\xi})(f(\mathbf{b}, \pmb{\xi}) - \mu(f(\mathbf{b}, \pmb{\xi})))^2 d{\pmb{\xi}}.
    \end{split}
\label{eq:mean_variance}
\end{equation}

Instead of Eq.~(\ref{eq:F}), we can also use one of $P(f(\mathbf{b}, \pmb{\xi}))$'s quantiles as $F(\cdot)$. The 100\% quantile represents the ``worst-case" scenario, while other less conservative quantile orders are also employed~\cite{baudoui2012local}.

On the other hand, RBDO refers to the optimization scheme where reliability analysis is incorporated into deterministic optimization methods~\cite{choi2007reliability}. Herein, reliability is defined as the probability that a system is expected to successfully operate under risks of interest, such as deflection, leakage, and local damage. RBDO approaches exploit stochastic methods to address the statistical nature of constraints and design problems. Given $m$ risk factors, a representative formulation of RBDO reads:
\begin{equation}
\begin{split}
\min_{\mathbf{b}} ~~~~& J_{\text{RBDO}}(\mathbf{b}, \mu_{\pmb{\xi}})=-f(\mathbf{b}, \mu_{\pmb{\xi}}) \\
\mathrm{s.t. } ~~~~& P_j\left[g_j(\mathbf{b}, \pmb{\xi})\geq0\right] \geq {R_j}, ~~~~j=1, \cdots, m,
\end{split}
\label{eq:RBDO}
\end{equation}
where $\mu_{\pmb{\xi}}$ is the mean of $ \pmb{\xi} $, $ g_j(\cdot) $ denotes the $j$-th limit-state function that indicates the margin of safety with respect to the $j$-th risk factor, and ${R_j}$ is the specified reliability level with respect to the $j$-th factor. Given the $j$-th factor, $g_j(\cdot)<0$ and $g_j(\cdot)\geq0$ denote the associated failure region and safe region, respectively.

\begin{figure}[t]
\centering
\includegraphics[width=0.6\textwidth]{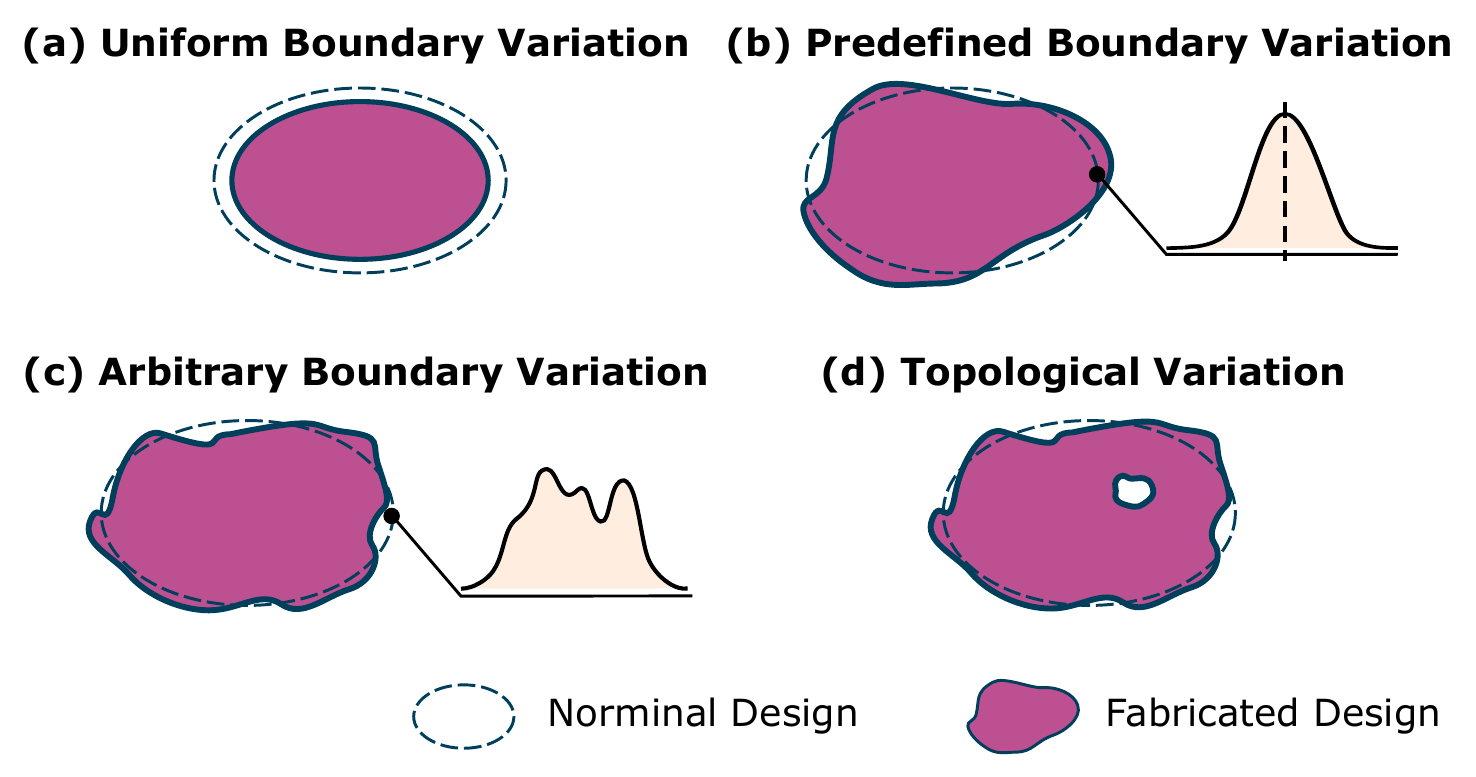}
\caption{Types of geometric uncertainty modeling: (a)~Uniform boundary variation where the boundary of the geometry is uniformly ``eroded" (\eg, over-etched) or ``dilated" (\eg, under-etched)~\cite{wang2019robust,da2019stress,sigmund2009manufacturing}; (b)~Predefined boundary variation where the distribution or the correlation of boundary points is predefined~\cite{chen2011new}; (c)~Arbitrary boundary variation where no assumption is imposed on the distribution of boundary points; (d)~Topological variation where the design's topological change (\eg, hole nucleation) is also possible. To the best of our knowledge, past work only considers (a) and (b) when modeling geometric uncertainty, while our proposed method can address ``free-form" uncertainties that include all four cases.}
\label{fig:geometric_uncertainty}
\end{figure}

Both approaches have been developed for design optimization under geometric uncertainty at various levels of geometric complexity (\ie, size, shape, and topology). Unlike parametric uncertainty that only considers the uncertainties of a few independent design parameters representing the design~\cite{morris2018design,wiest2022robust}, geometric uncertainty can account for the uncertainties at every point on the boundary of a geometry as well as its topological changes. 
% Among them, topology optimization under geometric uncertainty has been regarded as highly challenging due to modeling of topological uncertainty, propagation thereof, and stochastic design sensitivity analysis~\cite{chen2011new}. 
Due to the high-dimensionality of variables to be considered in geometric uncertainty quantification, previous work either assumes uniform boundary variation~\cite{wang2019robust,da2019stress,sigmund2009manufacturing} (Fig.~\ref{fig:geometric_uncertainty}a) or imposes predefined distribution/correlation on boundary points~\cite{chen2011new,lazarov2012topology,lazarov2012topology1,keshavarzzadeh2017topology,kang2018reliability} (Fig.~\ref{fig:geometric_uncertainty}b).
% or use spatially varying threshold field to mimic directionally random manufacturing tolerances~\cite{keshavarzzadeh2017topology}.
Specifically, \cite{chen2011new} models geometric uncertainty by perturbing the boundary using a random normal velocity field; \cite{lazarov2012topology,lazarov2012topology1,keshavarzzadeh2017topology,kang2018reliability} model geometric uncertainty by applying a predefined random field as the threshold of a level-set design representation; \cite{huang2014statistical,huang2015optimal,sabbaghi2018bayesian,ferreira2019automated} build shape deviation models and use Bayesian approaches to estimate the posterior of model parameters.
While those methods can simplify geometric uncertainty quantification by making assumptions on the form of shape deviations, the modeled uncertainties do not necessarily conform to realistic scenarios, which usually involve much more complicated geometric variability. For example, in real applications, the boundary variation does not necessarily follow canonical distributions (Fig.s~\ref{fig:geometric_uncertainty}c) and manufacturing defects do not only happen on the boundary~\cite{pham2021additive} (Fig.~\ref{fig:geometric_uncertainty}d). Therefore, how to model ``real-world", ``free-from" geometric uncertainty without making any simplifying assumptions is still an open challenge.

In this work, we overcome this challenge by using a hierarchical deep generative model to learn 1)~the underlying distribution of free-form nominal designs and 2)~the conditional distribution of fabricated design given any nominal design, under free-form geometric uncertainty (Figures~\ref{fig:geometric_uncertainty}c and \ref{fig:geometric_uncertainty}d). We demonstrate the efficacy using two real-world design examples. The ability to model free-form geometry and uncertainties allows us to address topological uncertainties.

%%%%%%%%%%%%%%%%%%%%%%%%%%%%%%%%%%%%%%%%%%%%%%%%%%%%%%%%%%%%%%%%%%%%%%
\section{METHODOLOGY}
\label{sec:methodology}

\begin{figure*}[t]
\centering
\includegraphics[width=1\textwidth]{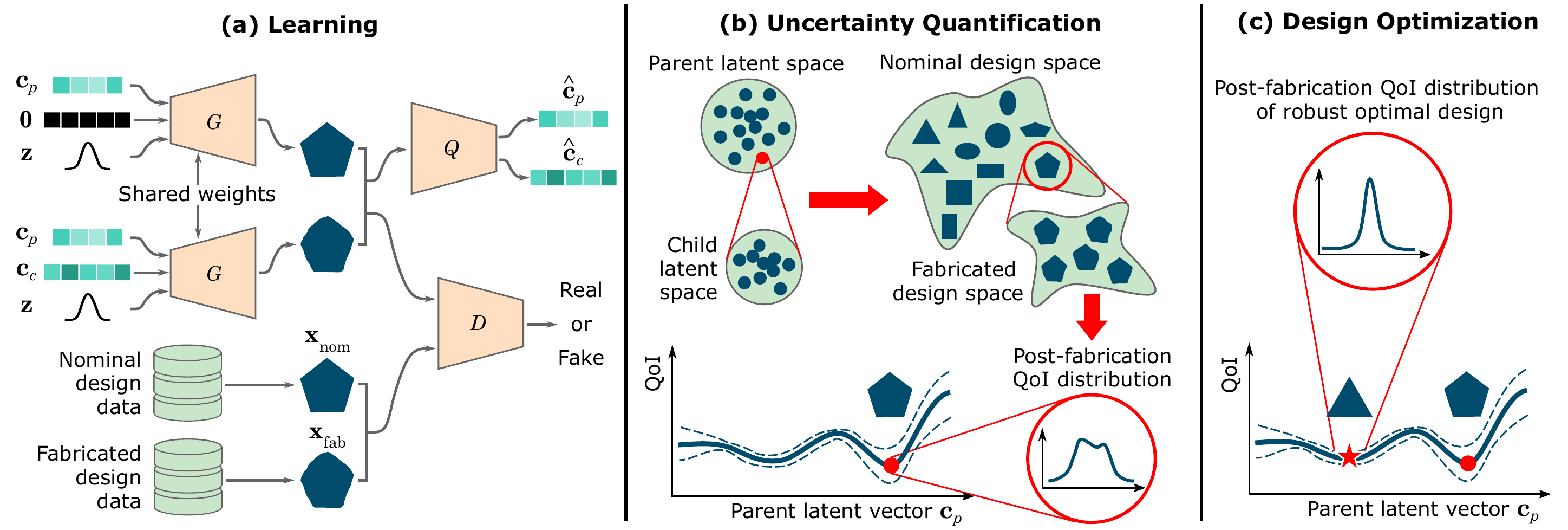}
\caption{Illustration of proposed Generative Adversarial Network-based Design under Uncertainty Framework (GAN-DUF): (a)~The proposed Hierarchical GAN architecture for simultaneously learning the compact representation of nominal designs and the conditional distributions of fabricated designs; (b)~Fabricated designs can be generated by sampling $\mathbf{c}_c$ at any fixed $\mathbf{c}_p$ representing a nominal design. The uncertainty of a nominal design's QoIs can be quantified by evaluating the QoIs of these generated fabricated designs via simulation or experiments; (c)~We can optimize $\mathbf{c}_p$ to obtain a nominal design $\mathbf{x}_{\text{nom}}$ with desired post-fabrication QoIs.}
% \caption{Illustration of proposed Generative Adversarial Network-based Design under Uncertainty Framework (GAN-DUF).}
\label{fig:architecture}
\end{figure*}

In this section, we introduce our Generative Adversarial Network-based Design under Uncertainty Framework (GAN-DUF). It consists of two parts: 1)~A generative adversarial network to learn a hierarchical latent representation that compactly captures both the variability of nominal designs and the uncertainty of any given nominal design\footnote{Code and data are available at \url{https://github.com/wchen459/GAN-DUF}.} (Sec.~\ref{sec:method_gan}); 2)~A design under uncertainty module that uses the trained generator to perform robust design optimization or reliability-based design optimization (Sec.~\ref{sec:method_opt}).

\subsection{Quantifying Uncertainty using Generative Adversarial Networks}
\label{sec:method_gan}

Let $\mathcal{I}_{\text{nom}}$ and $\mathcal{I}_{\text{fab}}$ denote the datasets of nominal and fabricated designs, respectively:
\begin{equation*}
\begin{split}
\mathcal{I}_{\text{nom}} &= \left\{\mathbf{x}_{\text{nom}}^{(1)},...,\mathbf{x}_{\text{nom}}^{(N)}\right\} \\
\mathcal{I}_{\text{fab}} &= \left\{\left(\mathbf{x}_{\text{fab}}^{(1,1)},...,\mathbf{x}_{\text{fab}}^{(1,M)}\right),...,\left(\mathbf{x}_{\text{fab}}^{(N,1)},...,\mathbf{x}_{\text{fab}}^{(N,M)}\right)\right\},
\end{split}
\end{equation*}
where $\mathbf{x}_{\text{fab}}^{(i,j)}$ is the $j$-th realization (fabrication) of the $i$-th nominal design. The \textbf{goals} are to 1)~learn a lower-dimensional, compact representation $\mathbf{c}$ of nominal designs to allow accelerated design optimization and 2)~learn the conditional distribution $P(\mathbf{x}_{\text{fab}}|\mathbf{c})$ to allow the quantification of manufacturing uncertainty at any given nominal design (represented by $\mathbf{c}$).

To achieve these two goals, we propose a generative adversarial network (Fig.~\ref{fig:architecture}a) that enables the hierarchical modeling of nominal designs and fabricated designs. Its generator $G$ generates fabricated designs when feeding in the parent latent vector $\mathbf{c}_p$, the child latent vector $\mathbf{c}_c$, and the noise $\mathbf{z}$; whereas it generates nominal designs simply by using the same generator $G$ but setting $\mathbf{c}_c=\mathbf{0}$. By doing this, we can control the generated nominal designs through $\mathbf{c}_p$ and the generated fabricated designs through $\mathbf{c}_c$. Note that the noise $\mathbf{z}$ represents the portion of geometric variation that cannot be captured by either the parent latent vector or the child latent vector. This is useful to be included as one of the inputs to the generator because sometimes the data contain noise (\eg, noisy or blurred images of fabricated samples) that is irrelevant to either the variation of nominal geometries or the fabrication imperfection. When using the trained model to generate nominal or fabricated designs, $\mathbf{z}$ is set to $\mathbf{0}$ to eliminate the effect of noise.

Given the pair of a generated nominal design $G(\mathbf{c}_p,\mathbf{0},\mathbf{z})$ and a generated fabricated design $G(\mathbf{c}_p,\mathbf{c}_c,\mathbf{z})$, the discriminator $D$ predicts whether the pair is generated or drawn from data (\ie, $\mathcal{I}_{\text{nom}}$ and $\mathcal{I}_{\text{fab}}$). Similar to InfoGAN, we also predict the conditional distribution $Q(\mathbf{c}_p, \mathbf{c}_c|\mathbf{x}_{\text{nom}}, \mathbf{x}_{\text{fab}})$ to promote disentanglement of latent spaces and ensure the latent spaces capture major geometric variability~\cite{chen2020airfoil}. The GAN is trained using the following loss function:
\begin{equation}
\begin{split}
\min_{G,Q}\max_D \mathbb{E}_{\mathbf{x}_{\text{nom}},\mathbf{x}_{\text{fab}}}[\log D(\mathbf{x}_{\text{nom}},\mathbf{x}_{\text{fab}})] + \\
\mathbb{E}_{\mathbf{c}_p,\mathbf{c}_c,\mathbf{z}}[\log(1-D(G(\mathbf{c}_p,\mathbf{0},\mathbf{z}),G(\mathbf{c}_p,\mathbf{c}_c,\mathbf{z})))] - \\ 
\lambda \mathbb{E}_{\mathbf{c}_p,\mathbf{c}_c,\mathbf{z}}[\log Q(\mathbf{c}_p,\mathbf{c}_c|G(\mathbf{c}_p,\mathbf{0},\mathbf{z}),G(\mathbf{c}_p,\mathbf{c}_c,\mathbf{z}))].
\end{split}
\end{equation}
As a result, $G$ decouples the variability of the nominal and the fabricated designs by using $\mathbf{c}_p$ to represent the nominal design (\textbf{Goal 1}) and $\mathbf{c}_c$ to represent the fabricated design of any nominal design. By fixing $\mathbf{c}_p$ and sampling from the prior distribution of $\mathbf{c}_c$, we can produce the conditional distribution $P(\mathbf{x}_{\text{fab}}|\mathbf{c}_p)=P(G(\mathbf{c}_p,\mathbf{c}_c,\mathbf{z})|\mathbf{c}_p)$ (\textbf{Goal 2}). As a neural network, the generator has the flexibility of generating fabricated designs completely based on real data, without making any unrealistic assumptions on the types of uncertainties.

Compared to existing uncertainty quantification methods, this GAN-based model opens up possibilities of 1)~building a universal UQ model compatible with both shape and topological designs, 2)~modeling \textit{free-form} geometric uncertainties without the need to make any assumptions on the distribution of geometric variability, and 3)~allowing fast prediction of uncertainties for new nominal designs.

The trained generator allows us to sample fabricated designs given any nominal design, simply by sampling the low-dimensional $\mathbf{c}_c$ with a fixed $\mathbf{c}_p$ representing the nominal design (Fig.~\ref{fig:architecture}b). We can then evaluate the quantities of interest (QoIs) of these generated fabricated designs using computational (\eg, physics simulation such as finite element methods, wave analysis, and computational fluid dynamics) or experimental methods. The QoIs may include performance, quality, properties, and/or cost. The resulted QoI distribution (\ie, \textit{post-fabrication QoI distribution}) allows us to quantify the uncertainty of QoIs for the nominal design. Note that the proposed framework is agnostic to both the type of designs (\eg, how designs are represented or what geometric variability is presented) and downstream tasks like design evaluation and design optimization.

\subsection{Design Optimization under Uncertainty using Trained Generator}
\label{sec:method_opt}

We can integrate the evaluated uncertainty into optimization frameworks such as robust optimization, where we simultaneously optimize the mean QoIs and minimize the influence of uncertainty~\cite{chen1996procedure} (Fig.~\ref{fig:architecture}c), as well as reliability-based optimization, where we optimize the QoIs subject to constraints such as the probability of failure or reliability index~\cite{choi2007reliability}. The solution is expected to maintain high real-world performance/quality, desired properties, or a lower chance of failure even under fabrication imperfection.

Specifically, in robust design optimization, the formulation of Eq.~(\ref{eq:RDO}) is modified as 
\begin{equation}
    \min_{\mathbf{c}_p} -F (\mu(f(\mathbf{x})), \sigma(f(\mathbf{x}))),
\end{equation}
where $\mathbf{x} = G(\mathbf{c}_p,\mathbf{c}_c,\mathbf{0})$. Here $\mathbf{c}_p$ and $\mathbf{c}_c$ corresponds to the vector of deterministic design variables and random variables, respectively.
In reliability-based design optimization, the formulation of Eq.~(\ref{eq:RBDO}) is modified as
\begin{equation}
\begin{split}
\min_{\mathbf{c}_p} ~& -f(G(\mathbf{c}_p,\mathbf{0},\mathbf{0})) \\
\mathrm{s.t. } ~& P_j\left[g_j(\mathbf{x})\geq0\right] \geq {R_j},~~~~j=1, \cdots, m,
\end{split}
\end{equation}
where $g_j(\mathbf{x})\geq0$ denotes that the $j$-th QoI is within an acceptable range. For example, the QoI can be the performance deviation from a given target performance.

In both formulations, we only need to optimize $\mathbf{c}_p$, which has a much lower dimensionality than the original design representation. The functionality of $\mathbf{c}_c$ is to introduce stochasticity and it reflects the sources of uncertainty.

%%%%%%%%%%%%%%%%%%%%%%%%%%%%%%%%%%%%%%%%%%%%%%%%%%%%%%%%%%%%%%%%%%%%%%
\section{RESULTS}
\label{sec:results}

We use two real-world robust design examples to demonstrate the effectiveness of our proposed framework. Ideally, to obtain fabricated design data $\mathcal{I}_{\text{fab}}$, we can take the nominal designs from $\mathcal{I}_{\text{nom}}$, fabricate them, and use the actual fabricated designs as data. However, in this study, we simulate the fabrication effects by deforming the geometry of nominal designs based on the following approaches, as a way to save time and cost. Note that how well the simulated manufacturing uncertainty resembles the real-world uncertainty is not central to this proof of concept study. We treat the simulated uncertainty as the real uncertainty only to demonstrate our design under uncertainty framework. In the ideal scenario, we can directly use the real-world fabricated designs to build $\mathcal{I}_{\text{fab}}$ and our proposed framework can still model the fabricated design distribution, since the framework is agnostic to the form of uncertainty. Also note that the required amount of data and latent vector dimensions will depend on the complexity level of geometric variation in data. For example, if the fabricated designs have a higher variation, we may need more fabricated design data and a higher-dimensional child latent vector to maintain the same level of accuracy for modeling the uncertainty. 

\subsection{Case Study: Airfoil Design} 

An airfoil is the cross-sectional shape of an airplane wing or a propeller/rotor/turbine blade. The shape of the airfoil determines the aerodynamic performance of a wing or a blade. By optimizing the airfoil shapes, we can improve their aerodynamic performance, minimize fuel consumption, and reduce greenhouse gas emissions.

\subsubsection{Dataset Construction}

We use the UIUC airfoil database\footnote{\url{http://m-selig.ae.illinois.edu/ads/coord_database.html}} as our nominal design dataset $\mathcal{I}_{\text{nom}}$. The preprocessing of $\mathcal{I}_{\text{nom}}$ and the creation of the fabricated design dataset $\mathcal{I}_{\text{fab}}$ are described as follows:

\paragraph{Nominal design data.} The original UIUC database contains invalid airfoil shapes and the number of surface coordinates representing each airfoil is inconsistent. Therefore, we used the preprocessed data from Chen \etal~\cite{chen2020airfoil} so that outliers are removed and each airfoil is consistently represented by 192 surface points (\ie, $\mathbf{x}_{\text{nom}}\in \mathbb{R}^{192\times 2}$). 

\begin{figure}[t]
\centering
\includegraphics[width=0.6\textwidth]{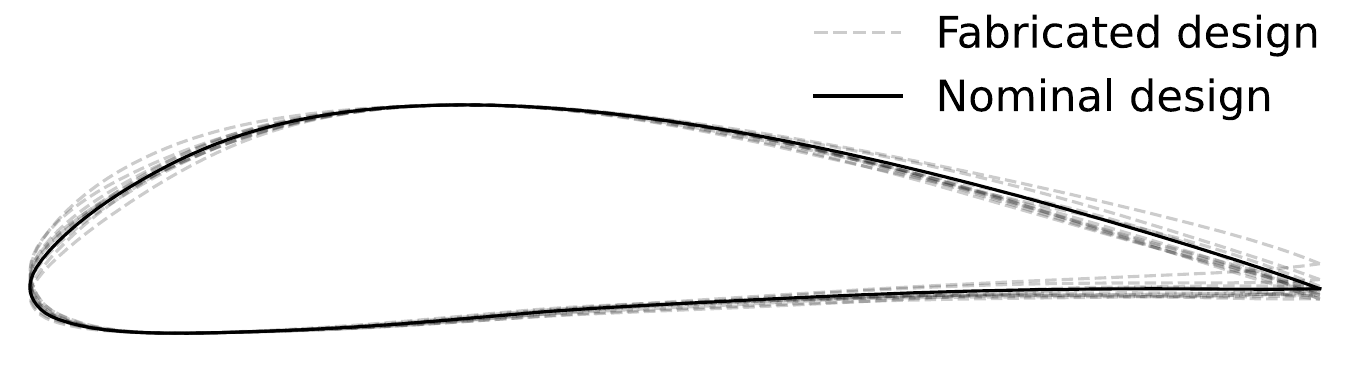}
\caption{An example of nominal airfoil design and its corresponding fabricated designs.}
\label{fig:airfoil_fabricated}
\end{figure}

\paragraph{Fabricated design data.} For airfoil designs, we simulate the effect of manufacturing uncertainty by randomly perturbing the free-form deformation (FFD) control points of each airfoil design from $\mathcal{I}_{\text{nom}}$~\cite{sederberg1986free}. Specifically, the original FFD control points fall on a $3\times 8$ grid and are computed as follows:
\begin{equation}
\begin{split}
& \mathbf{P}_{\text{nom}}^{l,m} = \\
& \left( x_{\text{nom}}^{\text{min}}+\frac{l}{7}(x_{\text{nom}}^{\text{max}}-x_{\text{nom}}^{\text{min}}), y_{\text{nom}}^{\text{min}}+\frac{m}{2}(y_{\text{nom}}^{\text{max}}-y_{\text{nom}}^{\text{min}}) \right), \\
& l=0,...,7 \text{ and } m=0,...,2,
\end{split}
\end{equation}
where $x_{\text{nom}}^{\text{min}}$, $x_{\text{nom}}^{\text{max}}$, $y_{\text{nom}}^{\text{min}}$, and $y_{\text{nom}}^{\text{max}}$ define the 2D minimum bounding box of the design $\mathbf{x}_{\text{nom}}$. To create fabricated designs, we add Gaussian noise $\epsilon\sim\mathcal{N}(0, 0.02)$ to the $y$-coordinates of control points except those at the left and the right ends. This results in a set of deformed control points $\{\mathbf{P}_{\text{fab}}^{l,m}|l=0,...,7;m=0,...,2\}$. The airfoil shape also deforms with the new control points and is considered as a fabricated design. The surface points of fabricated airfoils are expressed as
\begin{equation}
\mathbf{x}_{\text{fab}}(u,v)=\sum_{l=0}^{7}\sum_{m=0}^{2}B_l^7(u)B_m^2(v)\mathbf{P}_{\text{fab}}^{l,m},
\end{equation}
where $0\leq u\leq 1$ and $0\leq v\leq 1$ are parametric coordinates, and the $n$-degree Bernstein polynomials $B_i^n(u)=\binom{n}{i}u^i(1-u)^{n-i}$. We set the parametric coordinates based on the surface points of the nominal shape:
\begin{equation} 
(\mathbf{u}, \mathbf{v}) =
\left(
\frac{\mathbf{x}_{\mathrm{nom}}-x_{\mathrm{nom}}^{\mathrm{min}}}{x_{\mathrm{nom}}^{\mathrm{max}}-x_{\mathrm{nom}}^{\mathrm{min}}}, 
\frac{\mathbf{y}_{\mathrm{nom}}-y_{\mathrm{nom}}^{\mathrm{min}}}{y_{\mathrm{nom}}^{\mathrm{max}}-y_{\mathrm{nom}}^{\mathrm{min}}}
\right).
\end{equation}
Perturbing nominal designs via FFD ensures that the deformed (fabricated) shapes are still continuous, which conforms to conventional manufacturing methods for aerodynamic shapes. 
% By perturbing the FFD control points, we do not assume that the position of each point at the shape boundary follows a specific distribution, which justifies the need for a model that accounts for arbitrary boundary variation.
Figure~\ref{fig:airfoil_fabricated} shows an example of nominal design and its corresponding fabricated designs.

The final dataset contains 1,528 nominal designs and 10 fabricated designs per nominal design. Note that since similar nominal designs also have similar fabricated designs, we may need even fewer fabricated designs as training data. Please see Appendix~A for the study on how the sample size of fabricated design data affects uncertainty quantification using the proposed deep generative model.

\subsubsection{Generative Model Training and Evaluation}
\label{sec:airfoil_training}

We trained the proposed GAN on $\mathcal{I}_{\text{nom}}$ and $\mathcal{I}_{\text{fab}}$. We set the parent latent vector to have a uniform prior distribution $\mathcal{U}(\mathbf{0},\mathbf{1})$ (so that we can search in a bounded space during the design optimization stage), whereas both the child latent vector and the noise have normal prior distributions $\mathcal{N}(\mathbf{0},0.5\mathbf{I})$. The generator/discriminator architecture and the training configurations were set according to Chen \etal~\cite{chen2020airfoil}. During training, we set both the generator's and the discriminator's learning rate to 0.0001. We trained the model for 20,000 steps with a batch size of 32.

We conducted parametric studies over parent and child latent dimensions to investigate their effects on the generative performances (we fix the dimension of the noise $\mathbf{z}$ to 10)\footnote{Compared to parent and child latent dimensions, the noise dimension has a relatively small effect on the results as long as it is sufficiently large to capture the noise in training data~\cite{chen2020airfoil}.}. Particularly, we care about two performances: (1)~how well the parent latent representation can cover nominal designs, and (2)~how well the performance distributions of fabricated designs are approximated. The experimental settings and results are described as follows.

\begin{figure}[t]
\centering
\includegraphics[width=0.7\textwidth]{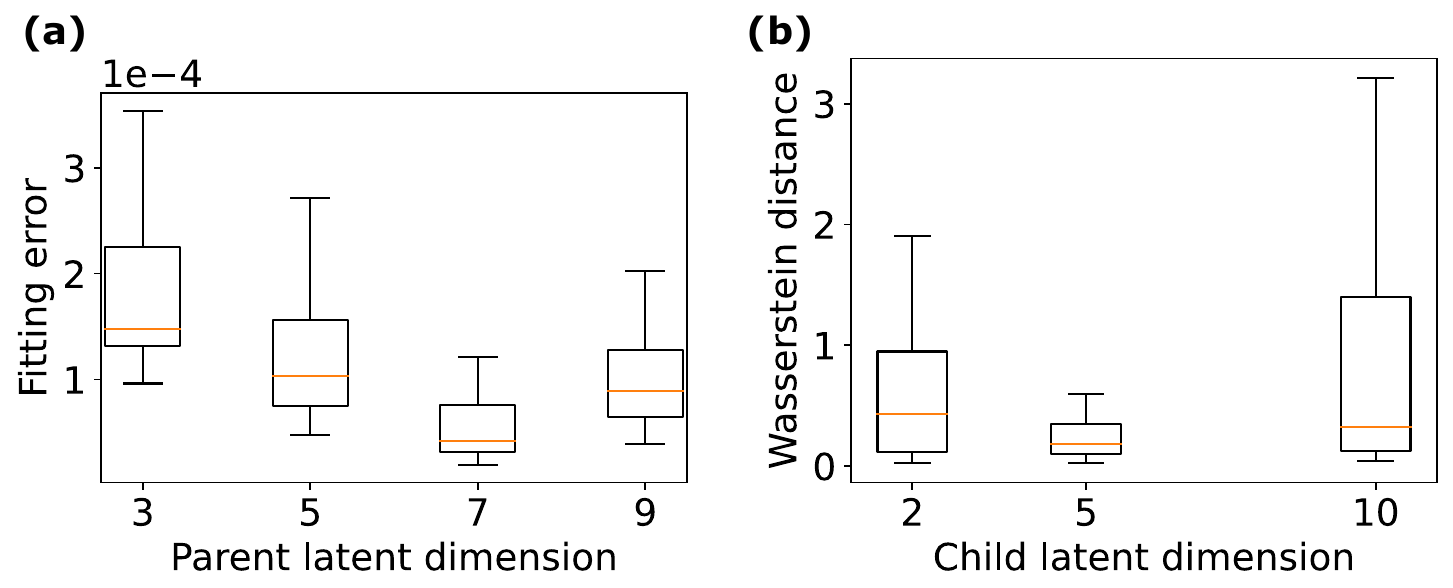}
% \caption{Parametric study showing (a)~how well the parent latent representation can cover nominal designs from data and (b)~how well the performance distributions of fabricated designs are approximated.}
\caption{Parametric study for the airfoil design example: (a)~The effect of parent latent dimensions on design space coverage indicated by fitting errors; (b)~The effect of child latent dimensions on conditional distribution approximation indicated by Wasserstein distances.}
\label{fig:parametric_study}
\end{figure}

We evaluated the first performance (\ie, nominal design coverage) via a fitting test, where we found the parent latent vector that minimizes the Euclidean distance between the generated nominal design and a target nominal design sampled from the dataset (\ie, fitting error). We use Sequential Least Squares Programming (SLSQP)~\cite{kraft1988software} as the optimizer and set the number of random restarts to 3 times the parent latent dimension. We repeated this fitting test for 100 randomly sampled target designs under each parent latent dimension setting. A parent latent representation with good coverage of the nominal design data will result in low fitting errors for most target designs. Figure~\ref{fig:parametric_study}a indicates that a parent latent dimension of 7 achieves relatively large design coverage (\ie, low fitting errors). 

We evaluated the second performance (\ie, fabricated design performance approximation) by measuring the Wasserstein distance between two conditional distributions~\textemdash~$P(f(\mathbf{x}_{\text{fab}})|\mathbf{x}_{\text{nom}})$ and $P(f(G(\mathbf{c}_p,\mathbf{c}_c,\mathbf{z}))|\mathbf{x}_{\text{nom}})$, where $f$ denotes the objective function. In this example, $f$ is the simulation that computes the lift-to-drag ratio $C_L/C_D$. For each generated nominal design $\mathbf{x}_{\text{nom}}$, we created 100 ``simulated" fabricated designs as $\mathbf{x}_{\text{fab}}$, the same way as we create training data, to be used as the ``ground truth" fabricated designs. Note that compared to the number of fabricated designs per nominal design in the training data, we created a much larger number of ``ground truth" fabricated designs for evaluation purposes. We also generated the same number of fabricated designs using the trained generator. We compute the aforementioned Wasserstein distance by using these two sets of samples. 
% Figure~\ref{fig:perf_distribution} shows a demonstrative example of a nominal design $\mathbf{x}_{\text{nom}}$ and the corresponding distributions $P(f(\mathbf{x}_{\text{fab}})|\mathbf{x}_{\text{nom}})$ and $P(f(G(\mathbf{c}_p,\mathbf{c}_c,\mathbf{z}))|\mathbf{x}_{\text{nom}})$. 
We repeated this test for 30 randomly generated nominal designs under each child latent dimension setting. Figure~\ref{fig:parametric_study}b shows that when the child latent dimension is 5, we have relatively low Wasserstein distances with the smallest variation (the parent latent dimension was fixed to 7). When the child latent dimension further increases to 10, the uncertainty of the Wasserstein distances increase, possibly due to the higher dimensionality. Note that the training data only contains 10 fabricated designs per nominal design, while in the test phase we use many more samples per nominal design to faithfully approximate the conditional distributions. We do not need that many samples in the training phase because the generative model does not learn independent conditional distributions for each nominal design, but can extract information across all nominal designs. 

\subsubsection{Design Optimization}

The objective of the design optimization is to maximize the lift-to-drag ratio $C_L/C_D$, which is computed by using the computational fluid dynamics (CFD) solver SU2~\cite{economon2016su2}.

We compared two optimization scenarios: 
\begin{enumerate}
\item Standard optimization, where we only consider the deterministic performance of the nominal design. The objective is expressed as $\max_{\mathbf{c}_p} f(G(\mathbf{c}_p,\mathbf{0},\mathbf{0}))$.
\item Robust design optimization, which accounts for the performance variation caused by manufacturing uncertainty. The objective is expressed as $\max_{\mathbf{c}_p} Q_{\tau} \left(f(G(\mathbf{c}_p,\mathbf{c}_c,\mathbf{0}))|\mathbf{c}_p\right)$,
where $Q_{\tau}$ denotes the conditional $\tau$-quantile. We set $\tau=0.05$ in this example.
\end{enumerate}

\begin{figure}[t]
\centering
\includegraphics[width=0.6\textwidth]{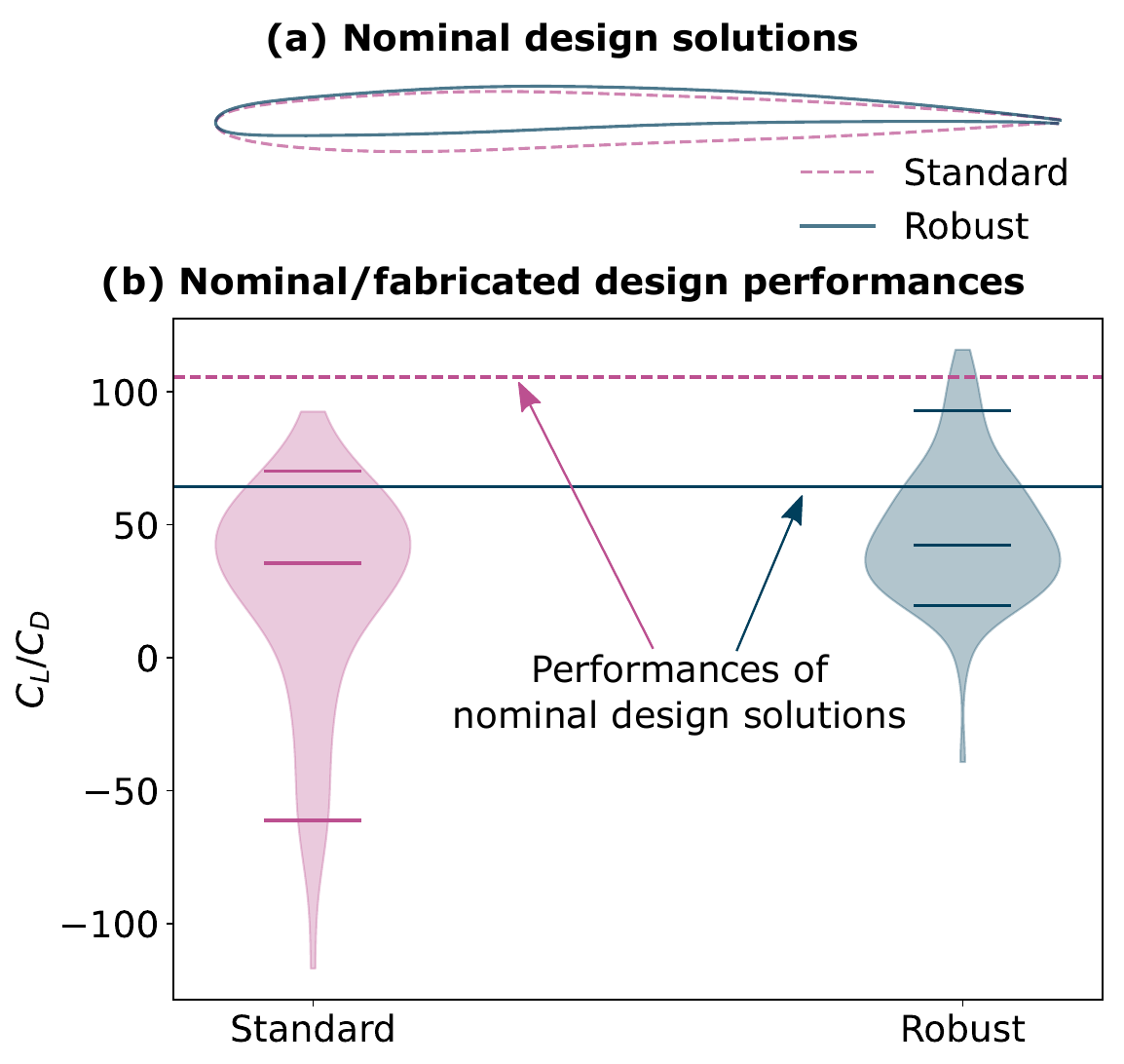}
\caption{Solutions for the airfoil design example: (a)~Nominal airfoil design solutions obtained by standard optimization and robust design optimization; (b)~When considering the manufacturing uncertainty, the robust design solution $\mathbf{x}^*_{\text{robust}}$ shows improved quantile values for the post-fabrication performance distribution compared to the standard design solution $\mathbf{x}^*_{\text{std}}$, even though the nominal performance of $\mathbf{x}^*_{\text{robust}}$ is slightly worse than $\mathbf{x}^*_{\text{std}}$. The short horizontal lines indicate 95\% quantiles, medians, and 5\% quantiles.}
% \caption{Solutions for the airfoil design example.}
\label{fig:opt_perf_distribution}
\end{figure}

Under both scenarios, we fixed the parent and the child latent dimensions to 7 and 5, respectively, based on the aforementioned parametric study. 
In each scenario, we performed Bayesian optimization (BO) to optimize $\mathbf{c}_p$. We evaluate 21 initial samples of $\mathbf{c}_p$ selected by Latin hypercube sampling (LHS)~\cite{mckay2000comparison} and 119 sequentially selected samples based on BO's acquisition function of expected improvement (EI)~\cite{jones1998efficient}\footnote{The settings of the initial and the total evaluation times in BO are based on the parent latent dimension $d_p$. Specifically, we performed $3d_p$ initial LHS evaluations and $20d_p$ total evaluations, where $d_p=7$ as mentioned earlier.}. In standard optimization, we evaluate the nominal design performance $f(G(\mathbf{c}_p,\mathbf{0},\mathbf{0}))$ at each sampled point. In robust design optimization, we estimate the quantile of fabricated design performances $f(G(\mathbf{c}_p,\mathbf{c}_c,\mathbf{0}))$ by Monte Carlo (MC) sampling using 100 randomly sampled $\mathbf{c}_c\sim P(\mathbf{c}_c)$ at each $\mathbf{c}_p$. Figure~\ref{fig:opt_perf_distribution} shows the design solutions and the distributions of ground-truth fabricated design performances\footnote{``Ground-truth fabricated design" refers to designs created by the same means by which the designs from $\mathcal{I}_{\text{fab}}$ were created.} (\ie, post-fabrication performance distributions) of these solutions. 
We also performed permutation tests~\cite{fisher1936design} to evaluate the difference between the post-fabrication performances of the standard and the robust design solutions (Table~\ref{tab:pvalue}).
The results show that, by accounting for manufacturing uncertainty, the post-fabrication performances of the robust design solution $\mathbf{x}^*_{\text{robust}}$ are significantly improved (indicated by small $p$-values), compared to the standard design solution $\mathbf{x}^*_{\text{std}}$, even though the nominal performance of $\mathbf{x}^*_{\text{robust}}$ is worse than $\mathbf{x}^*_{\text{std}}$. This result illustrates the possibility that the solution discovered by standard optimization can have high nominal performance but is likely to possess low performance when it is fabricated. The robust design optimization enabled by GAN-DUF can substantially mitigate this risk.

\begin{table}[t]
\caption{Statistical significance $p$-values obtained by permutation tests on the post-fabrication performances of the standard and the robust airfoil design solutions.  The test statistics are the improvement of the quantiles and the mean. A lower $p$-value indicates a more significant improvement of the robust design solution over the standard design solution in terms of the post-fabrication performances.}
\begin{center}
\label{tab:pvalue}
\begin{tabular}{c l}
\hline
Measures of location & $p$-value \\
\hline
5\% quantile & 1e-5 \\
25\% quantile & 8e-5 \\
Median & 0.0397 \\
Mean & 1e-5 \\
\hline
\end{tabular}
\end{center}
\end{table}

\subsection{Case Study: Optical Metasurface Absorber Design}

\begin{figure}[t]
\centering
\includegraphics[width=0.6\textwidth]{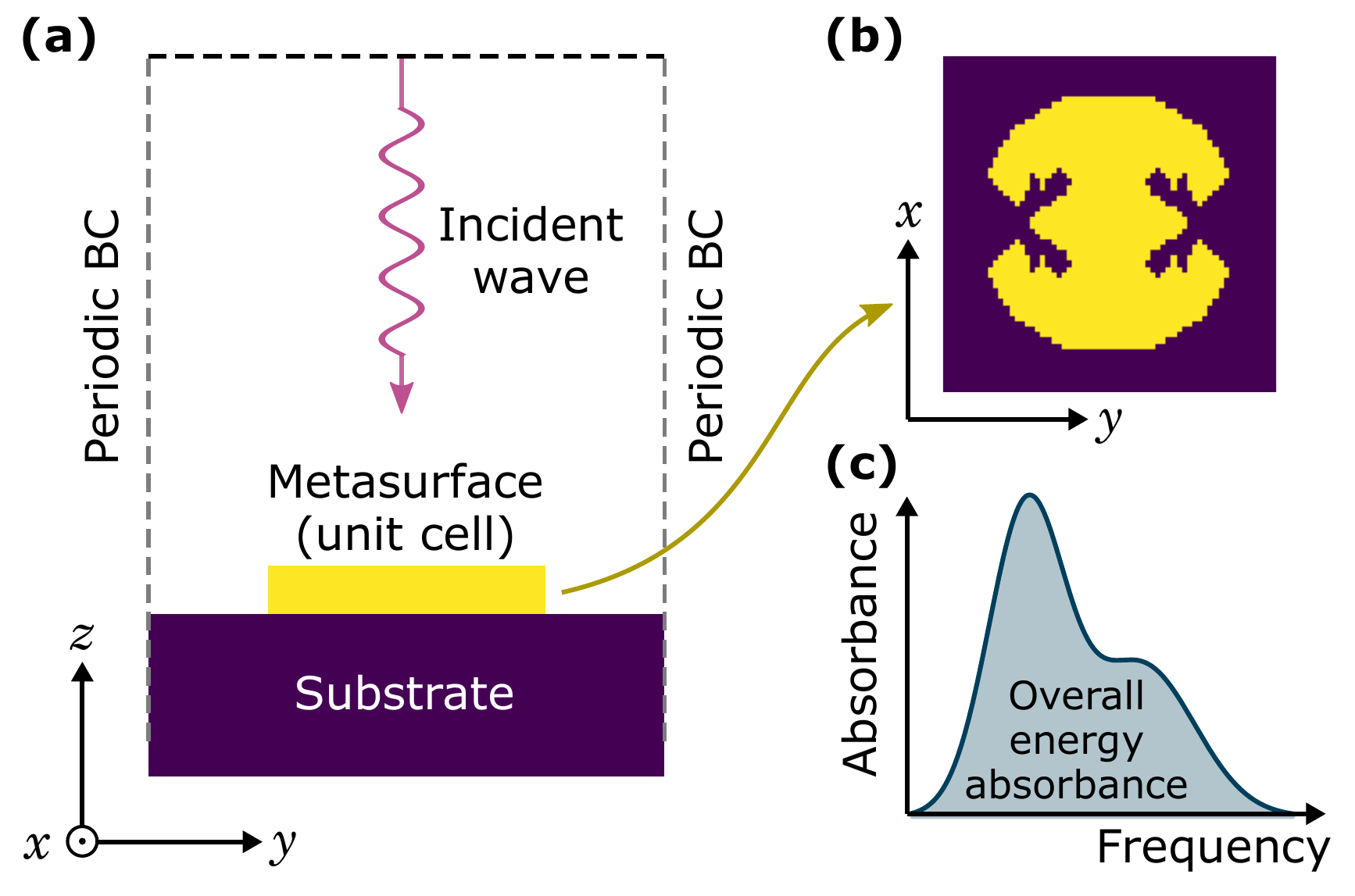}
% \caption{Solutions for the metasurface design example.}
\caption{Conceptual illustration of a metasurface absorber: (a) A schematic of wave analysis. A unit cell is placed on top of the dielectric substrate. A plane wave is normally incident to the metasurface. Periodic boundary conditions are imposed on the lateral surfaces of the analysis domain; (b) An example of a free-form unit cell (nominal design); (c) A hypothetical frequency response of absorbance $A$ at a target frequency band.}
\label{fig:wave_analysis}
\end{figure}

Optical metasurfaces are artificially engineered structures that can support exotic light propagation building on subwavelength inclusions~\cite{chen2016review, bukhari2019metasurfaces}. Among many, metasurface absorbers~\cite{liu2017experimental} have been intensely studied for engineering applications such as medical imaging, sensing, and wireless communications. In this work, the functionality of interest is large energy absorbance at a range of incident frequencies in the terahertz regime. Fig.~\ref{fig:wave_analysis} shows a schematic of wave analysis. A free-form unit cell made of Au is placed on top of a dielectric substrate whose relative permittivity is set as 2.88-0.09$i$, where the imaginary term is associated with energy loss. The periodic boundary condition on electromagnetic fields is imposed on the lateral surfaces of the analysis domain. The energy absorbance performance at a single frequency $\omega$ is quantified as $A(\omega)=1-|S_{11}(\omega)|^2$, where $\omega$ is the excitation frequency of the incident wave, and $S_{ij}$ is the component of the $S$-parameter matrix that characterizes an electrical field intensity from port $i$ to port $j$ in a complex network. This absorbance behavior largely depends on the cross-sectional topology of the unit cell and, by extension, is subject to deviation under fabrication uncertainty.

\subsubsection{Dataset Construction}
\label{sec:metasurface_data}

\begin{figure}[t]
\centering
\includegraphics[width=0.6\textwidth]{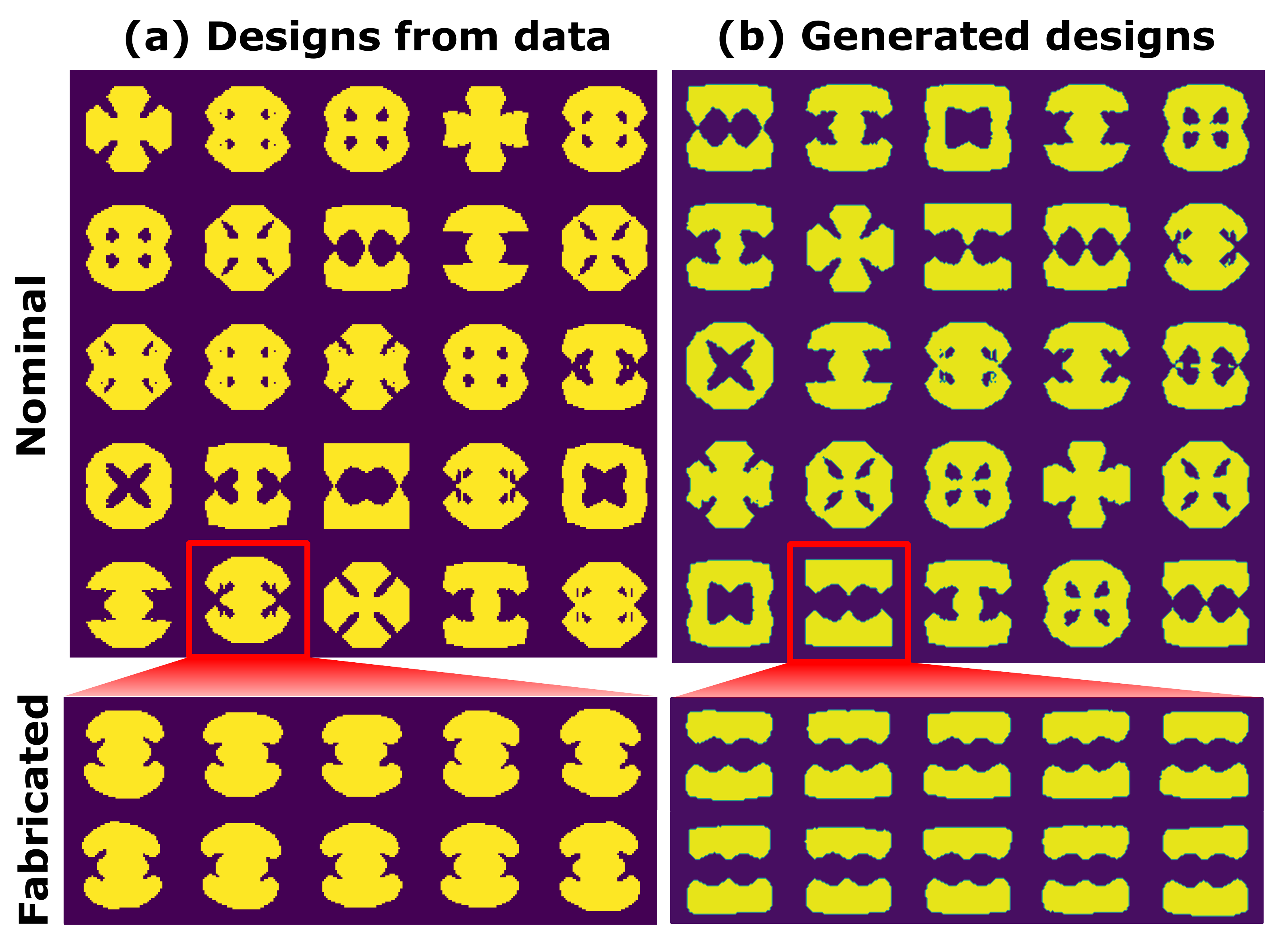}
\caption{Visual inspection on generated designs: (a)~Metasurface designs randomly drawn from training data; (b)~Designs randomly generated from a trained generator.}
\label{fig:metasurface_samples}
\end{figure}

We created 1,000 nominal designs and 10 fabricated designs per nominal design (Fig.~\ref{fig:metasurface_samples}a) by using the following method:

\paragraph{Nominal design data.} The nominal design dataset builds on three topological motifs (\ie, I-beam, cross, and square ring)~\cite{larouche2012infrared, azad2016metasurface}. We create nominal designs by randomly interpolating the signed distance fields of these baselines~\cite{whiting2020meta}. As a result, each design is stored as $64\times 64$ level-set values (\ie, $\mathbf{x}_{\text{nom}}\in \mathbb{R}^{64\times 64}$). We can obtain final designs by thresholding the signed distance fields. Building on a given set of baselines, this shape generation scheme allows a unit cell population that is topologically diverse.

\paragraph{Fabricated design data.} Similar to the airfoil design example, we randomly perturb a set of $12\times 12$ FFD control points in both $x$ and $y$ directions with white Gaussian noise that has a standard deviation of 1 pixel. This leads to the distortion of the $64\times 64$ grid coordinates at all the pixels, together with the level-set value at each pixel. We then interpolate a new signed distance field as the fabricated (distorted) design.
To account for the limited precision of fabrication, we further apply a Gaussian filter with a standard deviation of 2 to smooth out sharp, non-manufacturable features. 
% Similar to the airfoil example, by perturbing the FFD control points, we do not assume that the position of each point at the shape boundary follows a specific distribution, which justifies the need for a model that accounts for arbitrary boundary variation. 
The Gaussian filter can also change the topology by, \eg, removing small holes.
% , which justifies the need for a model that accounts for topological variation. 
To demonstrate that our proposed deep generative model is flexible in addressing more complicated uncertainties, we also construct a scenario where fabrication errors contain non-smooth boundary variation and random hole nucleation. We show that our proposed model can also address this more complicated scenario. Please refer to Appendix~B for details.

\subsubsection{Generative Model Training and Evaluation}
\label{sec:metasurface_training}

\begin{figure}[t]
\centering
\includegraphics[width=0.5\textwidth]{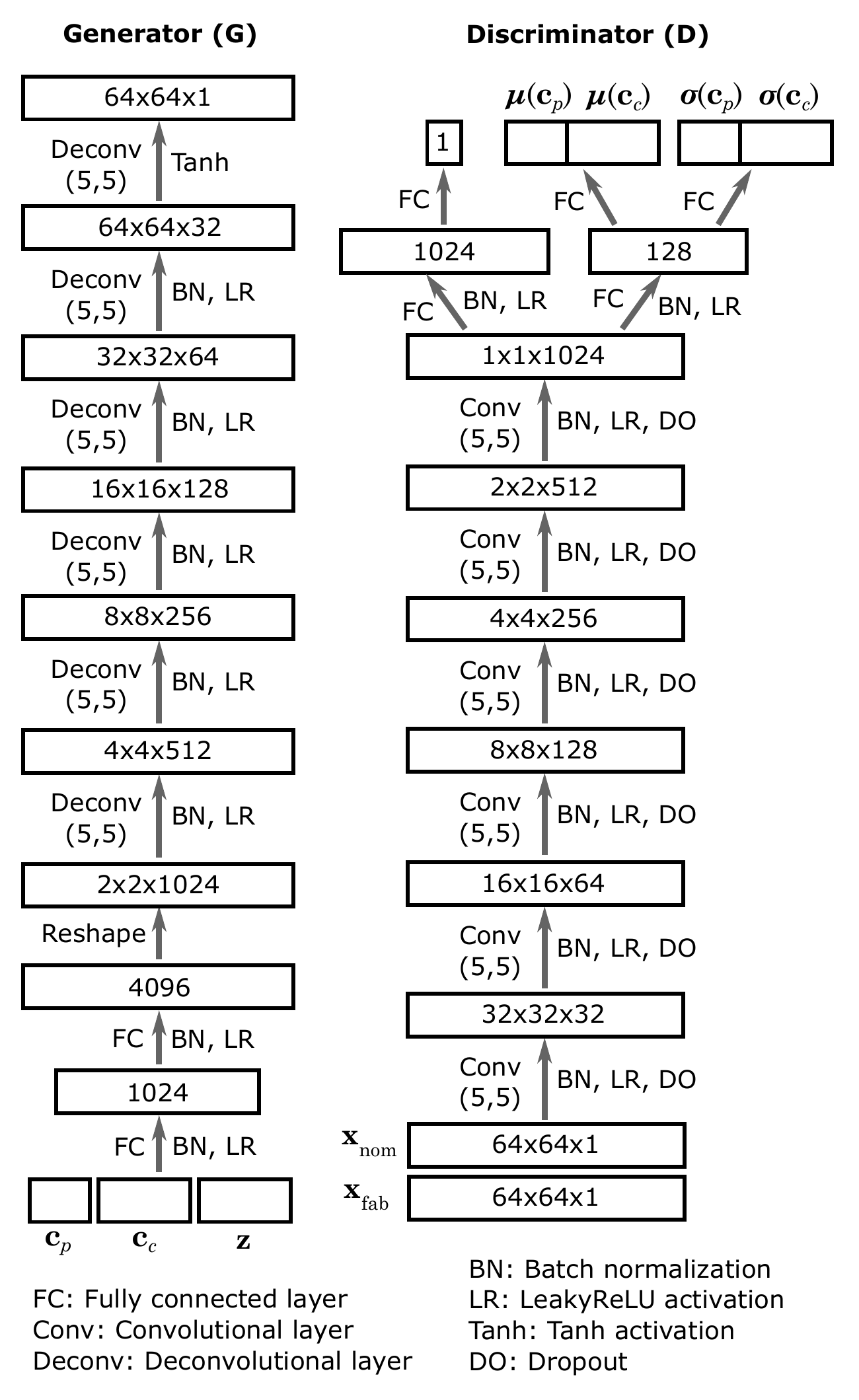}
\caption{Generator and discriminator architectures in the metasurface design example.}
\label{fig:metasurface_configuration}
\end{figure}

As mentioned in the Background section, optimizing designs with varying topology under geometric uncertainty hosts a great challenge. GAN-DUF can handle this problem by modeling the uncertainty using the proposed generative adversarial network. Same as the airfoil example, we set the parent latent vector to have a uniform prior distribution, while both the child latent vector and the noise have normal prior distributions. Again, we fixed the noise dimension to 10. The generator and the discriminator architectures are shown in Fig.~\ref{fig:metasurface_configuration}. The discriminator predicts both the discriminative distribution $D(\mathbf{x}_{\text{nom}},\mathbf{x}_{\text{fab}})$ and the auxiliary distribution $Q(\mathbf{c}_p,\mathbf{c}_c|\mathbf{x}_{\text{nom}},\mathbf{x}_{\text{fab}})$. During training, we set both the generator's and the discriminator's learning rate to 0.0001. We trained the model for 50,000 steps with a batch size of 32.

\begin{figure}[t]
\centering
\includegraphics[width=0.7\textwidth]{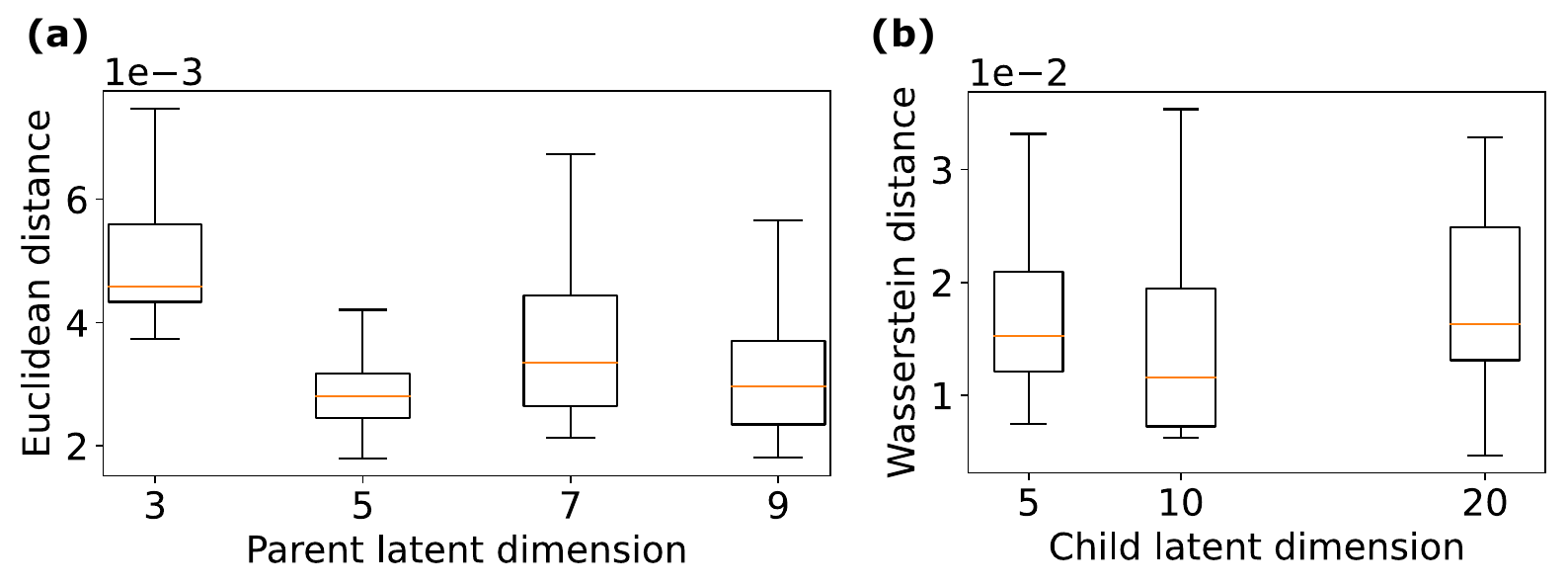}
\caption{Parametric study for the metasurface design example: (a)~The effect of parent latent dimensions on design space coverage indicated by fitting errors; (b)~The effect of child latent dimensions on conditional distribution approximation indicated by Wasserstein distances.}
\label{fig:metasurface_parametric_study}
\end{figure}

Similar to the airfoil design example, we want to investigate the effect of the parent and the child latent dimensions on the trained GAN's capability of covering the design space and approximating the fabricated design performance distributions. To evaluate the design space coverage, we performed a fitting test as described in Sec.~\ref{sec:airfoil_training}. Specifically, we use SLSQP as the optimizer and set the number of random restarts to 3 times the parent latent dimension. Here the fitting error is the Euclidean distance between the signed distance fields of the generated nominal design and a target nominal design sampled from the dataset. Under each parent latent dimension setting, we randomly select 100 target designs. Figure~\ref{fig:metasurface_parametric_study}a indicates that a parent latent dimension of 5 achieves sufficiently large design coverage, while further increasing the parent latent dimension cannot improve the coverage. 

To evaluate the trained GAN's performance on approximating the fabricated design performance distributions, we use the Wasserstein distance to measure the distributional difference between $P(f(\mathbf{x}_{\text{fab}})|\mathbf{x}_{\text{nom}})$ and $P(f(G(\mathbf{c}_p,\mathbf{c}_c,\mathbf{z}))|\mathbf{x}_{\text{nom}})$ (see Sec.~\ref{sec:airfoil_training}). Under each child latent dimension setting, we compute the Wasserstein distance between the energy absorbance distributions of ``ground truth" fabricated designs and generated fabricated designs. We performed fewer design evaluations than the airfoil design example due to the higher cost of wave analysis. Under each child latent dimension setting, we computed Wasserstein distances for 10 nominal designs, each with 30 ``ground truth" fabricated designs and 30 generated fabricated designs. Figure~\ref{fig:metasurface_parametric_study}b indicates that a child latent dimension of 10 achieves the lowest median and 25\%/75\% quantiles.

Figure~\ref{fig:metasurface_samples}b shows nominal and fabricated designs randomly generated from the trained generator with a parent and a child latent dimensions of 5 and 10, respectively.

\subsubsection{Design Optimization}

During the design optimization stage, we fixed the parent and the child latent dimensions to be 5 and 10, respectively, based on the aforementioned parametric study. The objective is to maximize the mean energy absorbance over the given range of frequencies, 8-9 THz. To achieve broadband functionality, we formulate the objective function as the sum of energy absorbance at individual frequencies (\ie, $f= \sum_{i=1}^{n_\omega} A(\omega_i)$, where $n_\omega$ is the number of frequencies at which absorbance is to be observed). The RF Module of COMSOL Multiphysics\textsuperscript{\textregistered}~\cite{multiphysics1998introduction} is used for evaluation of metasurfaces.

\begin{figure}[t]
\centering
\includegraphics[width=0.6\textwidth]{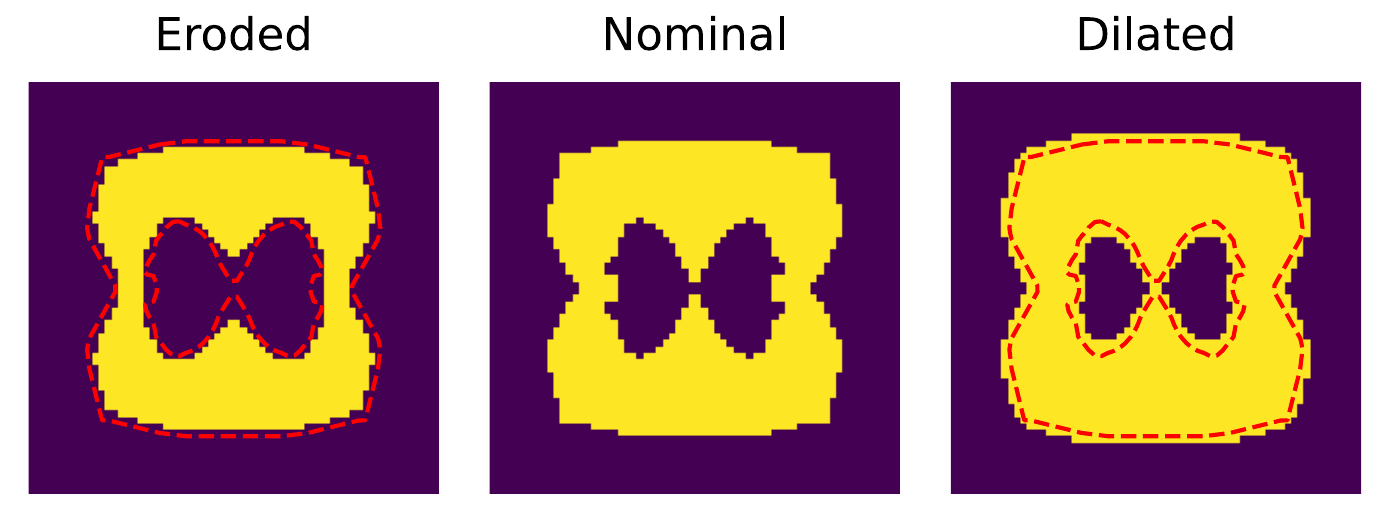}
% \caption{Solutions for the metasurface design example.}
\caption{An example of metasurface design with fabrication uncertainty modeled as uniform boundary variation~\cite{wang2019robust,da2019stress,sigmund2009manufacturing}.}
\label{fig:metasurface_dilated_eroded}
\end{figure}

Similar to the airfoil design example, we compared standard optimization with robust design optimization. To investigate the benefit of modeling free-form uncertainty over uniform uncertainty, we also benchmarked the performance of GAN-DUF against the method that models fabrication uncertainty as uniform boundary variation~\cite{wang2019robust,da2019stress,sigmund2009manufacturing}. Specifically, following \cite{wang2019robust}, we first convolve the nominal design pattern $\mathbf{x}_{\text{nom}}(r)$ with a Gaussian filter to obtain a blurred pattern $\tilde{\mathbf{x}}(r)$:
\begin{equation}
    \tilde{\mathbf{x}}(r) = \sum_{r_i}\frac{1}{\alpha}\mathbf{x}_{\text{nom}}(r_i)e^{-\frac{(r-r_i)^2}{\sigma^2}},
\end{equation}
where $r$ is the spatial coordinate on the pattern and $\alpha$ is a normalization factor defined as
\begin{equation}
    \alpha = \sum_{r_i}e^{-\frac{(r-r_i)^2}{\sigma^2}}.
\end{equation}
The coefficient $\sigma$ controls the blur radius that corresponds to patterning resolution. In our experiment, we set $\sigma = 2$. We simply threshold the blurred pattern $\tilde{\mathbf{x}}(r)$ to create eroded and dilated patterns $\bar{\mathbf{x}}(r)$:
\begin{equation}
\bar{\mathbf{x}}(r)=
    \begin{cases}
    0, 0 \leq \tilde{\mathbf{x}}(r) \leq \eta \\
    1, \eta \leq \tilde{\mathbf{x}}(r) \leq 1
    \end{cases},
\label{eq:threshold}
\end{equation}
where $\eta$ is the threshold with a value between 0 and 1. We can create an eroded pattern by setting $\eta>0.5$ and a dilated pattern by setting $\eta<0.5$. The eroded and dilated patterns represent extreme fabrication errors. The edge deviation $\Delta$ (\ie, the distance of boundary shift) is related to both $\eta$ and $\sigma$:
\begin{equation}
    \frac{1}{2} - \eta = \text{erf}\left\{\frac{\Delta}{\sigma}\right\}.
\label{eq:edge_deviation}
\end{equation}
In this experiment, we set $\Delta$ to -0.4 and 0.4 for the eroded and the dilated patterns, respectively. We can compute $\eta$ based on Eq.~(\ref{eq:edge_deviation}) and then obtain the eroded/dilated pattern using Eq.~\ref{eq:threshold}. Please see \cite{wang2019robust} for more details. Figure~\ref{fig:metasurface_dilated_eroded} shows an example of resulted patterns. While we manually select $\sigma$ and $\Delta$ as in \cite{wang2019robust}, we can also calibrate this parameter using training data. However, this calibration will require additional study and experimentation, and hence may deviate from our goal of comparing our method to existing ones. Nevertheless, considering data-driven parameter calibration for the existing methods could be an interesting future work.

To account for this uniform boundary variation as the uncertainty in robust design optimization, we set the objective to be $f(\mathbf{x}_{\text{nom}})+0.5f(\bar{\mathbf{x}}_{\text{eroded}})+0.5f(\bar{\mathbf{x}}_{\text{eroded}})$, where $\bar{\mathbf{x}}_{\text{eroded}}$ and $\bar{\mathbf{x}}_{\text{dilated}}$ denote the eroded and the dilated patterns, respectively, and $f$ computes the overall energy absorbance for a given metasurface pattern. Due to the prohibitive cost of optimizing a high-dimensional design ($\mathbf{x}_{\text{nom}}\in\mathbb{R}^{64\times 64}$) without analytical gradient, we perform the robust design optimization in the latent space instead, \ie, based on $\mathbf{x}_{\text{nom}}=G(\mathbf{c}_p,\mathbf{0},\mathbf{0})$. We will use ``robust (uniform)" to refer to this optimization scenario.

In each optimization scenario, we performed BO with 15 initial LHS samples and 85 sequentially selected samples based on the acquisition strategy of EI\footnote{We performed $3d_p$ initial LHS evaluations and $30d_p$ total evaluations in BO, where $d_p=5$ is the parent latent dimension.}. The quantile of fabricated design performances at each $\mathbf{c}_p$ was estimated from 30 MC samples. We used fewer MC samples than those in the airfoil design case due to the higher cost of evaluating the objective (\ie, performing wave analysis to compute the energy absorbance).

\begin{figure}[t]
\centering
\includegraphics[width=0.6\textwidth]{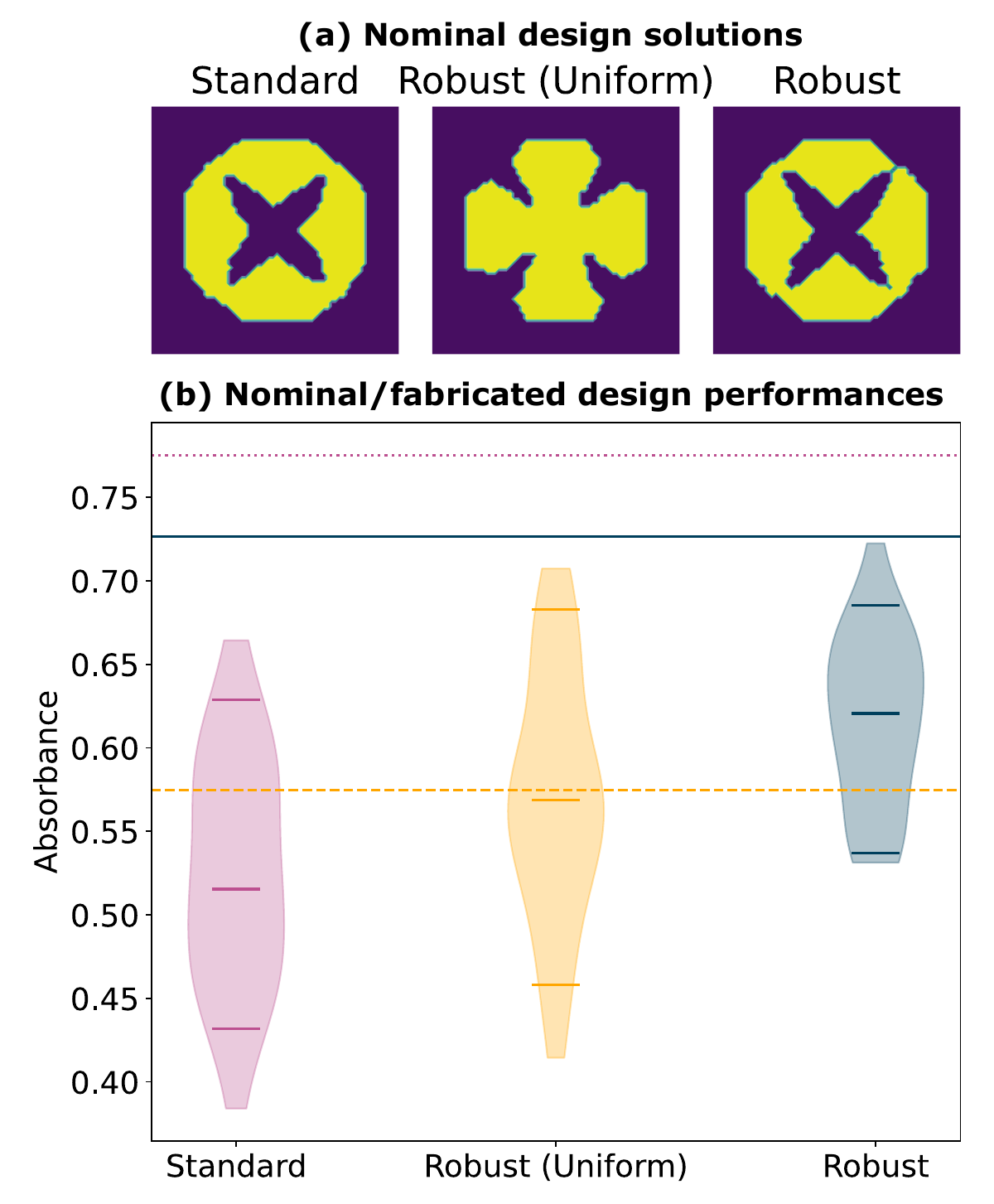}
% \caption{Solutions for the metasurface design example.}
\caption{Solutions for the metasurface design example: (a)~Nominal metasurface design solutions obtained by standard optimization and robust design optimization; (b)~When considering the manufacturing uncertainty, the robust design solution $\mathbf{x}^*_{\text{robust}}$ shows improved post-fabrication performance distribution compared to the standard design solution $\mathbf{x}^*_{\text{std}}$, even though the nominal performance of $\mathbf{x}^*_{\text{robust}}$ is worse than $\mathbf{x}^*_{\text{std}}$. The short horizontal lines indicate 95\% quantiles, medians, and 5\% quantiles.}
\label{fig:metasurface_opt_perf_distributions}
\end{figure}

Figure~\ref{fig:metasurface_opt_perf_distributions} shows the design solutions and the distributions of ground-truth fabricated design performances for these solutions. We observe similar trends as in the airfoil design case, where the standard optimization finds the solution with the highest nominal performance, while the two solutions obtained from robust optimization have higher performances (in general) after fabrication (Fig.~\ref{fig:metasurface_opt_perf_distributions}b). Particularly, by considering free-form uncertainty enabled by GAN-DUF, the robust design solution further improves upon the solution obtained under uniform uncertainty, in terms of their actual post-fabrication performances.
The low $p$-values in Table~\ref{tab:metasurface_pvalue} indicate the statistical significance of this improvement. This demonstrates that making simplifying assumptions on the form of uncertainty can lead to sub-optimal solutions. The proposed GAN-DUF does not suffer from this issue since it has the flexibility of learning and considering any form of uncertainty. 
This result is expected because the ability to accurately model the uncertainties is a prerequisite for finding the true robust design solution that performs better than others when considering uncertainties.

\begin{table}[t]
\caption{The $p$-values obtained by permutation tests that quantify the statistical significance of the improvement when considering free-form uncertainty enabled by GAN-DUF, in terms of the post-fabrication performances. The test statistics are the improvement of the quantiles and the mean. For example, a low value in row ``Median" and column ``Standard" indicates that there is a significant improvement of the free-form uncertainty-enabled robust design solution over the standard design solution in terms of the medians of their post-fabrication performance distributions.}
\begin{center}
\label{tab:metasurface_pvalue}
\begin{tabular}{c l l}
\hline
Measures of location & \multicolumn{2}{c}{$p$-value} \\
 & Standard & Robust (uniform) \\
\hline
5\% quantile & 1e-5 & 0.0031 \\
25\% quantile & 3e-5 & 0.00024 \\
Median & 1e-5 & 0.0118 \\
Mean & 1e-5 & 0.0028 \\
\hline
\end{tabular}
\end{center}
\end{table}

% \begin{table}[t]
% \caption{The $p$-values obtained by permutation tests to determine the statistical difference between the post-fabrication performances of the robust design solution and the solutions obtained under the other two optimization scenarios (\ie, standard design optimization and robust optimization with uniform boundary variation).}
% \begin{center}
% \label{tab:pvalue}
% \begin{tabular}{c l}
% \hline
% Measures of location & $p$-value \\
% \hline
% 5\% quantile & 1e-5 \\
% 25\% quantile & 3e-5 \\
% Median & 1.3e-4 \\
% Mean & 1e-5 \\
% \hline
% \end{tabular}
% \end{center}
% \end{table}

%%%%%%%%%%%%%%%%%%%%%%%%%%%%%%%%%%%%%%%%%%%%%%%%%%%%%%%%%%%%%%%%%%%%%%
\section{CONCLUSION}
\label{sec:conclusion}

We proposed GAN-DUF to facilitate design under free-form geometric uncertainty. It contains a novel deep generative model that simultaneously learns a compact representation of nominal designs and the conditional distribution of fabricated designs given any nominal design. The proposed framework is generalizable to any geometric design representations (\ie, both shape and topological designs) and can address free-form uncertainties without resorting to any simplifying assumption on the type of uncertainty. We applied GAN-DUF to two real-world engineering design examples (namely aerodynamic shape optimization and metasurface absorber topology optimization) and showed its capability in finding the design solution that is more likely to possess a better performance after fabrication or manufacturing. 

In this work, we assume that fabricated designs are independent. In the case where the distribution of fabricated designs depends on the varying fabrication tool condition due to, \eg, tool wear, we can introduce an additional latent vector $\mathbf{c}_t$ that encodes the variation of the fabrication tool. In this way, the proposed generative model can estimate the conditional distribution of fabricated designs given the tool condition. We can also continuously calibrate $\mathbf{c}_t$ using new fabrication data.

Although we only considered fabrication/manufacturing uncertainty when demonstrating the proposed framework, it is also applicable to other sources of geometric uncertainties such as those caused by operational wear or erosion. In addition to robust design optimization demonstrated in this work, we can also combine the proposed hierarchical generative model with reliability-based design optimization to find designs that are less likely to fail after fabrication or under operational wear/erosion. 

GAN-DUF is generalizable to 3D designs with certain adjustments to neural network architectures, on top of the 2D cases demonstrated in this work. For example, if the 3D designs are represented as point clouds, one can adopt the generator/discriminator architecture from \cite{shu20193d}. Since 3D data may contain more complex geometric variation, it is likely that we need more layers, higher latent dimensions, and a larger dataset to accommodate the extra complexity.

While in this work we used synthetic fabrication data, for future work, we will collect real fabricated designs as training and test data to validate the effectiveness of GAN-DUF in a completely realistic scenario.

%%%%%%%%%%%%%%%%%%%%%%%%%%%%%%%%%%%%%%%%%%%%%%%%%%%%%%%%%%%%%%%%%%%%%%
\section*{Acknowledgement}

This work was supported by the NSF CSSI program (Grant No. OAC 1835782) and the Northwestern McCormick Catalyst Award. We thank the anonymous reviewers for their comments.

%Bibliography
\bibliographystyle{unsrt}  
\bibliography{references}  

%%%%%%%%%%%%%%%%%%%%%%%%%%%%%%%%%%%%%%%%%%%%%%%%%%%%%%%%%%%%%%%%%%%%%%
\appendix       %%% starting appendix
\section*{Appendix A: A Sample Size Study}

\begin{figure}[t]
\centering
\includegraphics[width=0.6\textwidth]{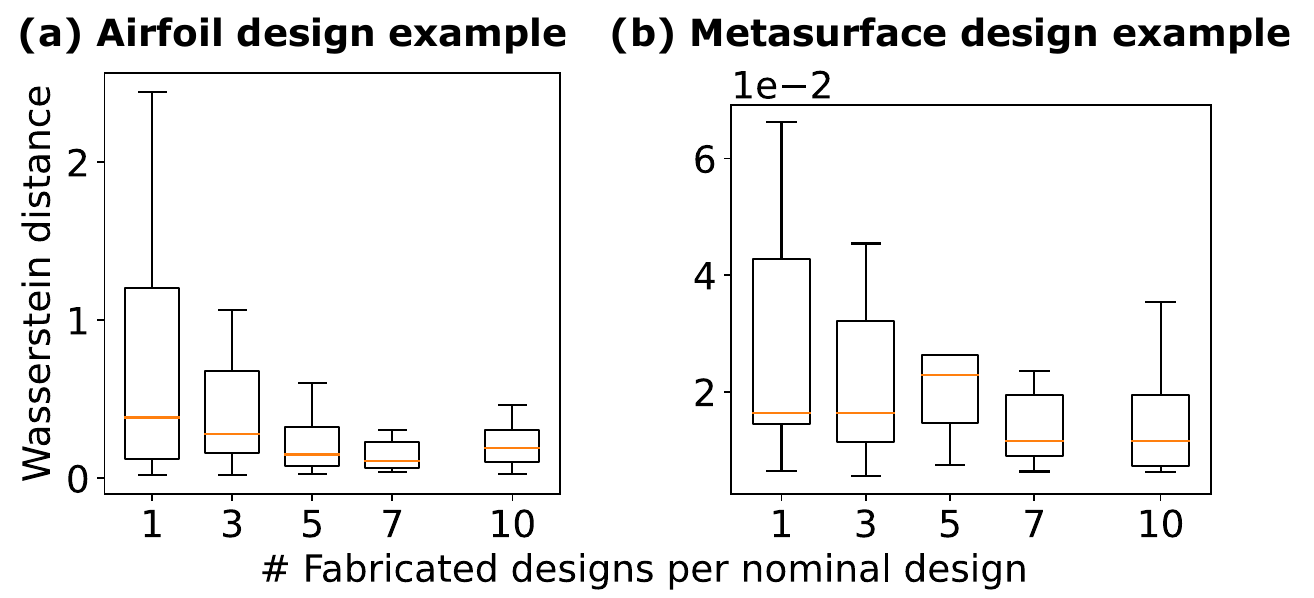}
\caption{Sample size study showing how the number of fabricated designs per nominal design affects the approximation of fabricated designs' performance distribution.}
\label{fig:sample_size}
\end{figure}

In the main text, we fixed the number of fabricated designs per nominal design in the training data. In practice, the sample size of fabricated designs may differ across use cases. there is a trade-off between the cost of acquiring fabrication data and the accuracy of quantifying fabrication uncertainties. To investigate how the sample size affects uncertainty quantification and draw insights into the minimum fabrication data requirement, we perform a sample size study on the fabrication data. We vary the number of fabricated designs per nominal design in the training data while fixing other hyperparameters. After training the deep generative model, we evaluate how well the performance distributions of fabricated designs are approximated by the trained generator. We measure this in the same way as the parametric study over child latent dimensions described in Sec.~\ref{sec:airfoil_training} and Sec.~\ref{sec:metasurface_training}~\textemdash~\ie, using the Wasserstein distance to measure the distributional difference between $P(f(\mathbf{x}_{\text{fab}})|\mathbf{x}_{\text{nom}})$ and $P(f(G(\mathbf{c}_p,\mathbf{c}_c,\mathbf{z}))|\mathbf{x}_{\text{nom}})$. Results in Fig.~\ref{fig:sample_size} show a general trend of decreased Wasserstein distance or improved performance distribution approximation, as the sample size of the fabricated designs increases. Among the tested cases, we observe the most accurate approximation, in both the airfoil and the metasurface design examples, when there are 7 fabricated designs per nominal design in the training data. Further increasing the number of fabricated designs per nominal design to 10 does not lead to a significant improvement in the approximation accuracy. Our future work will study how GAN-DUF will perform in design under uncertainty under sparse samples of fabricated designs.

%%%%%%%%%%%%%%%%%%%%%%%%%%%%%%%%%%%%%%%%%%%%%%%%%%%%%%%%%%%%%%%%%%%%%%
\appendix       %%% starting appendix
\section*{Appendix B: More Complicated Uncertainties}

Due to the flexibility of deep neural networks, our proposed GAN does not assume any specific forms of uncertainties. To help demonstrate this point, we create a scenario for the metasurface design example where fabrication uncertainties include defects of random hole nucleation and non-smooth boundary deviations. 

\begin{figure*}[t]
\centering
\includegraphics[width=0.6\textwidth]{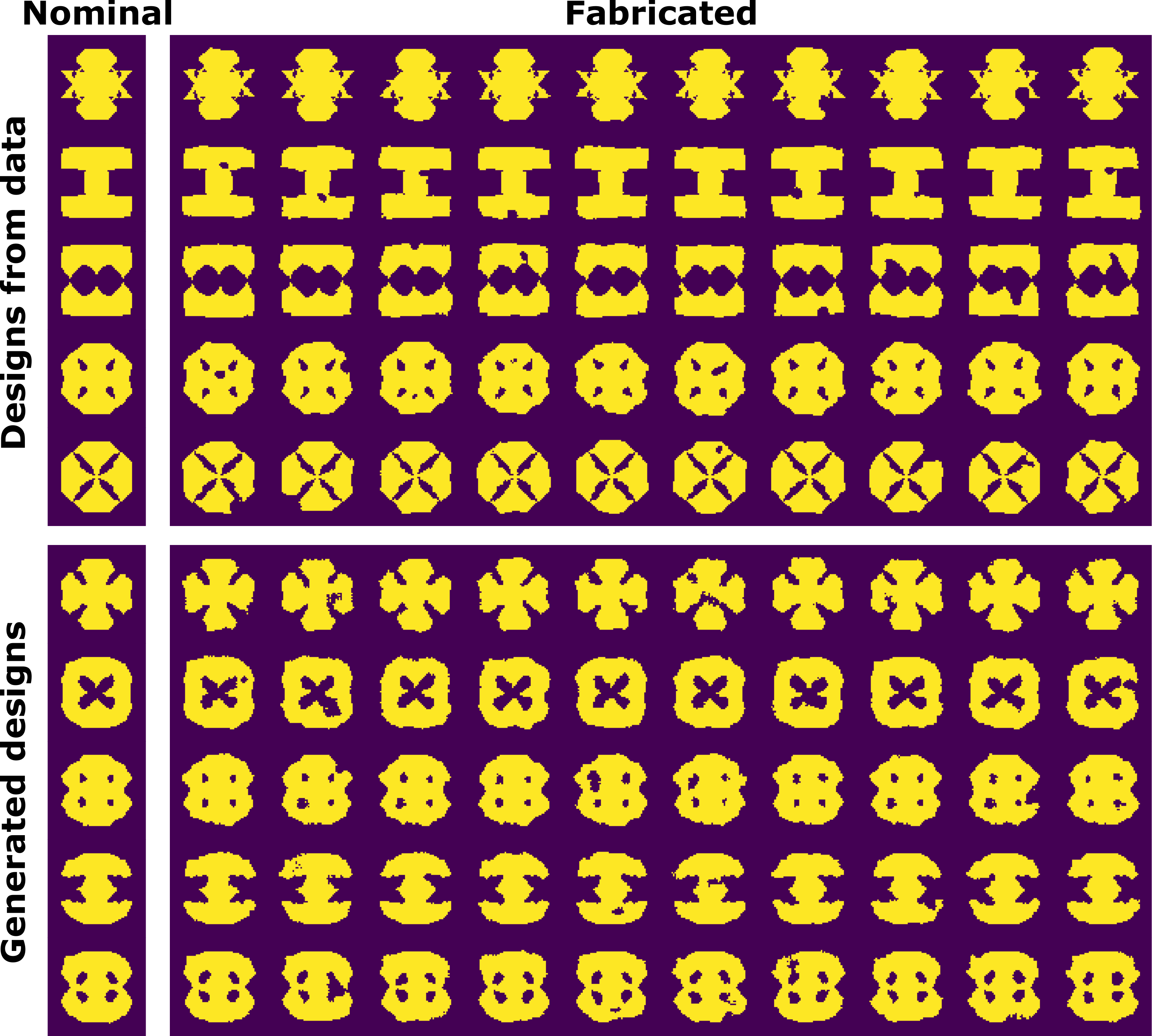}
\caption{The proposed GAN can address fabrication uncertainties with non-smooth shape variation and hole nucleation.}
\label{fig:hole_sharp}
\end{figure*}

Specifically, compared to the fabricated design data construction mentioned in Sec.~\ref{sec:metasurface_data}, we remove the Gaussian filter, increase the FFD lattice resolution (\ie, the number of FFD control points), and double the random perturbation scale of FFD control points, to create non-smooth boundary deviations. We also create a hole with a random shape and size at a random location of each fabricated design\footnote{The hole will not show if its location is sampled outside the metasurface pattern.}. The nominal design data remain unchanged. Examples of fabricated design data are shown in the upper part of Fig.~\ref{fig:hole_sharp}.

We show that by some minor adjustments of hyperparameters, our proposed GAN can capture these more complicated uncertainties, as shown in the lower part of Fig.~\ref{fig:hole_sharp}. To adapt to the more complicated fabrication data, we decrease the convolutional/deconvolutional kernel size to (3, 3), increase the child latent dimension to 20, and double the number of filters in the generator's deconvolutional layers. The visual comparison shows that GAN-DUF can generate fabricated designs with random hole nucleation and non-smooth boundary variation as occurred in the data, indicating its ability to capture these more complicated uncertainties.

\end{document}